# Origin of the Unusual Temperature Dependence of the Upper Critical Field of Kagome Superconductor CsV$_3$Sb$_5$: Multiple Bands or van Hove Singularities?

Ramakanta Chapai,[1, *, §] A. E. Koshelev,[2, †] M. P. Smylie,[1, 3] D. Y. Chung,[1] Asghar Kayani,[4] Khushi Bhatt,[5] Gaurab Rimal,[4] M. G. Kanatzidis,[1,6] W.-K. Kwok,[1] J. F. Mitchell,[1] and Ulrich Welp[1, ‡]

[1]*Materials Science Division, Argonne National Laboratory, Lemont, IL 60439, USA*
[2]*Department of Physics and Astronomy, University of Notre Dame, Notre Dame, IN 46656, USA*
[3]*Department of Physics and Astronomy, Hofstra University, Hempstead, NY 11549, USA*
[4]*Department of Physics, Western Michigan University, Kalamazoo, MI 49008, USA*
[5] *Physics Division, Argonne National Laboratory, Lemont, IL 60439, USA*
[6]*Department of Chemistry, Northwestern University, Evanston, IL 60201, USA*

Van Hove singularities (vHs) located close to the Fermi level in Kagome superconductors AV$_3$Sb$_5$ (A = K, Rb, Cs) have profound influence on their electronic and transport characteristics. Specifically, magneto-transport and susceptibility measurements on CsV$_3$Sb$_5$ reveal an anomalous temperature dependence of the upper critical field $H_{c2}(T)$, characterized by a pronounced upward curvature for both in-plane and *c*-axis magnetic fields, with zero-temperature $H_{c2}$ values of ~6.0 T and ~1.2 T, respectively. Our theoretical analysis, using a newly developed single-band model incorporating vHs and gap anisotropy, suggests that the observed upper critical field behavior is predominantly driven by the anisotropy of the Fermi velocity originating from vHs, instead of multi-band effects or gap anisotropy. Increased electron scattering introduced by proton irradiation defects smears out the vHs, reduces anisotropy, and recovers the conventional $H_{c2}(T)$ behavior, corroborating our proposed model.

* chapair@ornl.gov | † akoshelev@live.com | ‡ welp@anl.gov

§ Current address: Materials Science and Technology Division, Oak Ridge National Laboratory, Oak Ridge, TN 37831, USA

The appearance of van Hove singularities (vHs), Dirac cones, flat bands, and geometrical frustration are hallmarks of Kagome lattices, making this structural motif a fruitful platform for the exploration of frustrated magnetism, spin liquid states, electronic correlations, non-trivial band topology, anomalous Hall effect and various types of charge order [1-6]. Accordingly, the recent discovery of superconductivity ($T_c \sim$ 0.9-3 K) and its complex interplay with charge density wave (CDW) order ($T_{CDW} \sim$ 78-104 K) in the metallic Kagome lattice compounds $AV_3Sb_5$ ($A$ = K, Rb, and Cs) has generated substantial interest in this class of materials [7-17]. A distinctive characteristic of these compounds is the proximity of the Fermi energy to vHs, as revealed by density functional theory (DFT) calculations [8, 18-23], and as observed in angle-resolved photoemission spectroscopy (ARPES) [8, 21, 22, 24-26]. This proximity is a possible contributor driving the CDW transition [27-30]; it also induces enhanced electronic density of states in specific regions of the Fermi surface, accompanied by a high degree of *anisotropy* in the Fermi velocity. Such anisotropy manifests in the macroscopic properties of $AV_3Sb_5$. Indeed, we have recently shown [31] that the unconventional magneto-transport properties of $CsV_3Sb_5$ can be explained by considering the anisotropic Fermi surface structure. Other quantities expected to reflect this Fermi velocity anisotropy include the upper critical field, $H_{c2}$ [32]. However, disentangling the effects of Fermi surface anisotropy, multiple bands, and possible gap anisotropy on the superconducting properties has proven to be challenging [33].

In this letter, we present a combined experimental and theoretical study of the upper critical field of $CsV_3Sb_5$ and its anisotropy. Our purpose is to link the superconducting phase boundary, $H_{c2}(T)$, to the Fermi surface structure, and to the question of multiband superconductivity in this Kagome metal [5, 13, 16, 34-40]. Measurements of $H_{c2}(T)$ in single crystal $CsV_3Sb_5$ reveal a pronounced upward curvature for *c*-axis as well as *ab*-plane magnetic fields. By combining experimental data with theoretical modeling, we demonstrate that the observed upward curvature is driven by anisotropy of the Fermi velocity, rather than by multiband effects or gap anisotropy. In our theoretical framework, anisotropy of the Fermi velocity is induced by the proximity to vHs; the formalism also accounts for the gap anisotropy. We test this model via proton irradiation to induce defects that are expected to suppress the anisotropy and smear out the vHs [41-43]. Indeed, with increased proton dosing, the upwards curvature observed in $H_{c2}(T)$ in the pristine sample is progressively suppressed, ultimately recovering the conventional Maki-de Gennes dirty-limit [44-46] behavior expected of a single, isotropic gap and corroborating our model.

High-quality single crystals of CsV$_3$Sb$_5$ were grown and characterized as described in our earlier works [31, 47-49]. Electrical resistivity measurements were performed in a 9-1-1 T triple-axis vector magnet inside a dilution refrigerator (Bluefors LD400) using the standard four-probe method with current applied along the *ab*-plane. The pristine crystals display superconducting and CDW transitions near $T_c \sim 3.5$ K and $T_{CDW} \sim 94$ K, respectively, as shown in the Supplemental Materials Fig. S1 (a-b) [47]. Irradiation of the samples with 5-MeV protons was performed at the tandem van de Graaff accelerator at Western Michigan University [47, 50].

The upper critical field $H_{c2}(T)$ is deduced from the measurement of the temperature dependence of the electrical resistivity $\rho(T)$ in various fields applied along the *c*-axis and the *ab*-plane as presented in Figs. 1(a) and 1(b), respectively, for the pristine CsV$_3$Sb$_5$. Upon the application of a field, the superconducting transition shifts in an almost parallel fashion to lower temperatures; in high fields, it undergoes slight broadening. In the normal state, there is a noticeable magnetoresistance (*MR*) for $H \parallel c$, while the *MR* is significantly smaller for $H \parallel ab$. Such behavior is expected in a model of semi-classical magnetoresistance for an anisotropic layered material [51, 52]. From the data in Figs. 1(a-b), we construct the temperature-field superconducting phase diagram of pristine CsV$_3$Sb$_5$ (crystal S1), shown in Fig. 1(c), for both field orientations; we define $T_c$ as the midpoint of the transitions. The superconducting transitions display, a kink-like feature frequently seen on CsV$_3$Sb$_5$ crystals [8-10, 53-55]. Measurements on a second crystal yield identical results (shown in Fig. S2 [47]), indicating that the observed behavior is intrinsic. In particular, for both orientations, the $H_{c2}(T)$ phase boundary displays pronounced upwards curvature, see Fig. 1(c, d). This behavior cannot be accounted for in the standard Werthamer-Helfand-Hohenberg (WHH) description [44] and is often interpreted as a characteristic feature of superconductors possessing multiple gaps [13, 16, 35, 56-59] or an anisotropic gap [33, 60-62]. While Fig. 1(d) shows as-measured $H_{c2}$-data, in Fig. 1(c) data normalized to the Ginzburg-Landau (GL) upper critical fields $H_{c2,\alpha}^{GL} = -dH_{c2,\alpha}/dT \big|_{T_c} T_c$ are presented, highlighting that the upward curvature is more pronounced for the *c*-axis.

The resulting superconducting anisotropy, $\Gamma = H_{c2}^{ab}/H_{c2}^{c}$ exhibits a strong temperature dependence (see Fig. 2(d)). While a sizable superconducting anisotropy might be expected given the layered crystal structure, the clear decrease in $\Gamma(T)$ with decreasing temperature ($\approx 8.5$ near $T_c$ to $\approx 5.5$ at low temperatures) is not expected in the conventional single-band description. It is, however, another signature of either multi-gap or anisotropic gap superconductivity [33, 60-69].

In particular, the decrease in anisotropy with decreasing temperature is similar to that observed in FeAs-based superconductors [70] but opposite to that found for the archetypal multi-gap MgB2 [58, 71]. For a two-band superconductor, such temperature variation of the $H_{c2}$ anisotropy may arise when the band with the larger weight has a much smaller in-plane Fermi velocity and is less anisotropic [32]. The results shown in Fig. 1 (c, d) and Fig. 2(d) are in agreement with earlier reports [72] but do not show any signatures of an unexpected second phase transition at low temperatures that has been recently reported [73]. Measurements for different in-plane directions of the applied field yield $H_{c2}(T)$-curves similar to those shown in Fig. 1 with a weak 180-deg symmetry superimposed (see Fig. S6, Supplemental Materials [47]).

Kagome $A$V$_3$Sb$_5$ superconductors have a rather complicated electronic band structure [7], and a complete first-principles calculation of the upper critical field is currently not feasible. Therefore, we employ simplified models capable of providing insight into the data. As it is common to attribute a pronounced upward curvature of $H_{c2}(T)$ such as shown in Fig. 1(c, d) to multiple-band effects, we first explore a two-band model to describe our data. Our model assumes two open Fermi surface sheets in the form of warped cylinders, reminiscent of the Sb-derived sheet centered on the Γ–A line [20]. Each Fermi surface is characterized by a superconducting gap $\Delta_\alpha$, and an in-plane Fermi velocity, $\mathbf{v}_\alpha$, each assumed to be isotropic in the plane. Following previous work [32, 69], we can present the equation for the upper critical field in the following form [47]

$$-\ln t = \sum_\alpha w_\alpha \int_0^\infty \frac{t\,ds}{\sinh(ts)} \left[1 - \left\langle \exp\left(-\frac{\hbar^2 \mathbf{v}_\alpha^2 H}{8\pi k_B^2 T_c^2 \Phi_0} s^2\right)\right\rangle_\alpha\right]. \qquad (1)$$

This relation is based on the quasi-classical Eilenberger equations applicable to clean superconductors; paramagnetic limiting is not included. Here, $t \equiv T/T_c$ is the reduced temperature, $w_\alpha$ are the band weights defined as $w_\alpha = n_\alpha \Delta_\alpha^2 / \Sigma_\beta n_\beta \Delta_\beta^2$ and satisfying $\Sigma_\alpha w_\alpha = 1$ with the fractional density of states $n_\alpha = N_\alpha / \Sigma_\beta N_\beta$. For the case of a two-band system, one finds $w_1 = 1/(1 + r_n r_\Delta)$ and $w_2 = r_n r_\Delta/(1 + r_n r_\Delta)$, where $r_n = N_2/N_1$ and $r_\Delta = \Delta_2^2/\Delta_1^2$ are the ratios of the band density of states and of the squared gaps. In Eq. (1), $\langle ... \rangle_\alpha$ notates averaging over the $\alpha$ Fermi surface sheet, $H$ is the applied field, $k_B$ is the Boltzmann constant and $\Phi_0$ the flux quantum. In this model, the shape of $H_{c2}(T)$ is determined by *four* reduced parameters: the band weight $w_1$, the ratio of squares of in-plane Fermi velocities $r_v = v_{F2}^2/v_{F1}^2$, and two reduced hopping integrals: $t_{z1}/\varepsilon_{F1}$ and $t_{z2}/\varepsilon_{F2}$ [47]. In general, $H_{c2}(T)$ acquires an upward curvature when a band with larger weight has smaller Fermi velocity.

The simultaneous fitting of the in-plane and *c*-axis $H_{c2}(T)$ data yields curves that describe the data well, as shown in Fig. 1(c), with the fitting parameters included in the figure. This analysis yields zero-temperature $H_{c2}$ values of ~1.2 T (*c*-axis) and ~6.0 T (*ab*-plane). We note, however, that the fit depends on the product $r_n r_\Delta$ such that neither the gap ratio nor the individual gaps can be determined independently. The fit reveals that while the two band weights are approximately equal, the anisotropy of the Fermi velocities is very large, $r_v = 57.6$. A remarkable feature of the experimental results is the large enhancement of $H_{c2}$ over the GL value by more than a factor of two for the *c*-axis while the standard WHH model predicts a reduction to ~0.71 $H_{c2}^{GL}$. In the quasi-2D limit, that is, small *c*-axis hopping integrals, the zero-temperature enhancement resulting from our two-band model can be expressed in closed form as [47]: $H_{c2}(0)/H_{c2}^{GL} = 0.59\, r_v^{w_1}[w_1/r_v + (1-w_1)] \approx 2.1$. The key fit parameter required for this property is the large value of the Fermi velocity anisotropy $r_v\ (= 57.6)$.

Using the extremal cross sections of the Sb-derived Fermi surface sheet (warped cylinder) as predicted in DFT calculations [20], one finds a *c*-axis hopping integral of $t_z/E_F \sim 0.12$ for this sheet, which is remarkably close to the fitted value of 0.133 for the 'second' band. Then, in this assignment, it is the 'first' band that has a much smaller Fermi velocity resulting in the large ratio $r_v = 57.6$. The actual physical realization of the 'first' band is not specified in this model and the hopping integral $t_{z1}/E_F$ should be regarded merely as a quantification of *c*-axis dispersion. Nevertheless, in light of the highly anisotropic Fermi velocity, it is tempting to relate the 'first' band to the large hexagonal Fermi surface sheet in close proximity to the vHS [31].

Indeed, the presence of vHs in the electronic spectrum of Kagome superconductors naturally induces strong anisotropy of the Fermi velocity which can lead to a pronounced upward curvature of $H_{c2}(T)$, as demonstrated previously in the 2D case [74-76]. We therefore consider a single-band model in which the Fermi level is positioned close to the van Hove singularities and rewrite Eq. (1) for a single band with an anisotropic wavevector dependent Fermi velocity as given by the hyperbolic sections of the Fermi surface [31, 44, 47, 69], see also the inset of Fig. 1(d).

$$-\ln t = \int_0^\infty \frac{tds}{\sinh(ts)} \left\{ 1 - \left\langle \Omega^2 \exp\left[-\left(\frac{v_j^2}{\bar{v}_j^2} + \frac{v_k^2}{\bar{v}_k^2}\right) hs^2\right] \right\rangle \right\} \qquad (2)$$

Here, $h = \frac{H_{c2,i}(T)}{H_{i0}}$ is the reduced field with the field scale $H_{i0} = \frac{8\pi T_c^2 \Phi_0}{\hbar^2 \bar{v}_j \bar{v}_k}$, *i* denotes the direction of the applied field, and *j* and *k* are the directions orthogonal to *i*. $\bar{v}_i^2 = \langle v_i^2 \rangle$, and $\langle ... \rangle$ implies the averaging over the Fermi surface. For in-plane isotropic systems, $\bar{v}_x = \bar{v}_y$. In addition to the factors describing the gap anisotropy, this model is characterized by the following parameters: $r_m$, the ratio of the effective masses measured along and across the saddle-shaped dispersion, $\kappa_c = K_c/p_{u0}$, the ratio of the cut off vector $K_c$ and the separation from the van Hove point $p_{u0}$ (see Fig. S7 [47]), and $t_z/E_{vH}$, where $t_z$ is a hopping integral describing dispersion in the $c$-direction, and $E_{vH}$, the location of the van Hove point below the Fermi energy at $t_z = 0$. The weight $\Omega(\mathbf{k}_F)$ accounts for possible gap anisotropy, $\Delta(\mathbf{k}_F) \propto \Omega(\mathbf{k}_F)$ and is normalized by $\langle \Omega^2 \rangle = 1$. As described in more detail in the Supplemental Materials section (Eq. B23), we assume an in-plane gap anisotropy of the form $\Delta(p_v, p_u) = \Delta_0\sqrt{1 - c_u p_u^2/K_c^2 - c_v p_v^2/K_c^2}$ with maxima near the van Hove points as suggested by recent ARPES measurements [37]. Here, $p_u$ and $p_v$ represent the electron momenta along and across the vHs as shown in the inset of Fig. 1(d). $c_u$ and $c_v$ are constants that are constrained by the gap anisotropy $\Delta_{min}/\Delta_{max} \approx \sqrt{1 - c_u - r_m c_v}$.

We first consider effects of Fermi surface anisotropy only without any gap anisotropy. This version of our single-band model describes the $H_{c2}(T)$ data well, as shown in Fig. 1(d), with the parameters resulting from the simultaneous fit of the in-plane and out of plane data included in the figure. The specific equations underlying these fits are presented in the Supplementary Materials [47]. We note that the relative upward curvature of $H_{c2}(T)$ for the in-plane direction is noticeably weaker than for the $c$-axis direction. This feature is a hallmark of the strong influence of the van Hove singularity on the upper critical field [47]. Based on this model, the upward curvature of $H_{c2}(T)$ can be characterized by the ratio of the zero-temperature and GL extrapolated values: $\frac{H_{c2,z}(0)}{H_{GL,z}} \simeq 0.295 \frac{2w_z \kappa_c}{\ln(2w_z \kappa_c)}$ [47]. Here, $w_z = \sqrt{\varepsilon_{vH}/|t_z|}$, which applies when the Fermi level crosses the van Hove energy at a certain $c$-axis wave vector. Due to the strong inequality $\kappa_c \gg 1$, this result implies strong upward curvature of $H_{c2}(T)$, $H_{c2,z}(0) \gg H_{GL,z}$. Using the fit parameters from Fig. 1(d), we find $H_{c2}(T)/H_{GL,z} \approx 2.35$. The reason for this curvature is the strong variation of the Fermi velocity induced by the van Hove singularities.

The effect of gap anisotropy is illustrated with simulated $H_{c2}(T)$ curves shown in Fig. S9 [47]. We note that even a fairly large gap anisotropy of 5 ($\Delta_{max}/\Delta_{min}$) results in comparatively small changes in the temperature dependence of $H_{c2}$. This outcome reflects the fact that $H_{c2}$ is given in terms of averages over the Fermi surface sheets, see Eqs. (1) and (2). However, other quantities such as the specific heat or the penetration depth, which directly probe the excitation of quasiparticles across the gap, will be strongly affected by gap anisotropy [72, 77]. Interestingly, we find that the upwards curvature for the $c$-axis $H_{c2}$ is suppressed due to gap anisotropy. In order to regain a good fit to the data, we need to increase the value of $\kappa_c$, implying that the Fermi surface comes closer to the vHS and that anisotropy of the Fermi velocity around the Fermi surface increases. For instance, a gap anisotropy of $\Delta_{max}/\Delta_{min} \sim 5$, representative of the ARPES results in Ref. [37], requires a value of $\kappa_c \sim 11.3$ to fit our $H_{c2}(T)$-data (see Fig. S9, Supplementary Materials). Here, values of $\kappa_c$ of 10 and 11.3 would correspond to an anisotropy of the squared in-plane Fermi velocities (averaged over the $c$-axis dispersion) of 24 and 27, in reasonable agreement with the value of $r_v$ deduced from the two-band model. Thus, the central outcome of our analysis is that both approaches, muli-band/multi-gap as well as single-band vHs, require a large Fermi velocity anisotropy to account for the $H_{c2}$-data, while the detailed gap structure appears to be of secondary importance. This leads us to conclude that the large anisotropy of $v_F$, brought about by van Hove singularities located in close proximity to the Fermi surface, is the underlying physical cause for the observed $H_{c2}$-behavior.

The preceding analysis aligns with recent ARPES measurements [37] that reveal superconducting gaps on three Fermi surface sheets of pristine $CsV_3Sb_5$. The warped cylindrical sheet, originating from Sb orbitals centered on the Γ and A points, exhibits an isotropic gap with $\Delta \sim 0.40$ meV. Similarly, the triangular sheet centered at the K (H) points also shows an isotropic gap with $\Delta \sim 0.40$ meV. In contrast, the large hexagonal sheet around the Γ (A) points has a highly anisotropic gap, with values reaching nearly zero along the A-H direction and up to $\Delta \sim 0.65$ meV along the A-L direction. Despite this anisotropy, the average gap on this sheet appears also to be $\sim 0.40$ meV within the error bars. This gap structure is distinctly different from what has been observed for instance in the prototypical two-gap superconductor $MgB_2$ where the gap has a clear bimodal distribution with peaks centered around ~2 and ~7 meV corresponding to the π and σ-bands, respectively [78, 79]. In contrast, in $CsV_3Sb_5$ the gap distribution features a single peak around 0.4 meV riding on top of a broad base representing the gap anisotropy on the hexagonal Fermi surface sheet.

Further support for the description in terms of Fermi velocity anisotropy arises from ARPES experiments [36] on the doped compounds $Cs(V_{1-x}Ta_x)_3Sb_5$ and $Cs(V_{1-x}Nb_x)_3Sb_5$, which reveal for both materials isotropic gaps on all three Fermi surface sheets of ~0.8 meV and ~0.5 meV, respectively. Yet, the upwards curvature of $H_{c2}$(T) is still observed [80-82]. This finding is expected in our single-band description, as van Hove singularities near the Fermi surface are present in the doped samples [83, 84].

Scattering by disorder in the doped compounds may contribute to the homogenization of the gap structure while leaving the van Hove singularities intact. Here, we induce large numbers of defects via proton irradiation to tune the electronic anisotropy of $CsV_3Sb_5$ through enhanced electron scattering while maintaining superconductivity. The temperature dependence of the resistivity of sample S1 following irradiation to a dose of $6 \times 10^{16}$ p/cm$^2$ is shown in Fig. S1(e) [47]. Upon irradiation, the residual resistivity increases significantly, from $\rho_0 \approx 1.9$ μΩ cm (at 5 K) for pristine to $\rho_0 \approx 58$ μΩ cm for the irradiated sample. At the same time, $T_c$ decreases from 3.5 K (pristine) to 2.0 K (irradiated) and the anomaly associated with the CDW transition becomes unobservable (see Fig. 3, *End Matter*). Furthermore, the normal state *MR*, clearly seen for $H \parallel c$ in the pristine sample (Fig. 1(a-b)), is largely suppressed.

From the magneto-transport data presented in Figs. 2(a) and 2(b), superconducting phase diagrams are constructed for the irradiated sample as presented in Fig. 2(c). Note that upon irradiation the upwards curvature of $H_{c2}(T)$ seen in pristine samples is strongly suppressed and turns into a conventional dirty-limit Maki-de Gennes shape [44-46] while the temperature dependent superconducting anisotropy gives way to the standard temperature independent form as shown in Fig. 2(d). The upper critical field slopes at $T_c$ are -1.11 T/K and -0.18 T/K for *ab*-plane and *c*-axis respectively. From the extrapolated value of the critical fields, we estimate the GL coherence length in the irradiated sample to be $\xi_c(0) \approx 4\ nm$ and $\xi_{ab}(0) \approx 23\ nm$. This evolution of the superconducting phase boundary with increasing disorder is expected in a scenario of progressive smearing of the vHs by strong scattering (see *End Matter* for additional details).

In conclusion, we have investigated the temperature dependence of the upper critical field, $H_{c2}(T)$, and its anisotropy in $CsV_3Sb_5$ using both experimental and theoretical approaches. Our data reveal an anomalous temperature dependence of $H_{c2}(T)$, marked by a distinct upward curvature for both *c*-axis and in-plane magnetic fields, with zero-temperature values of ~1.2 T (*c*-axis) and ~6.0 T (*ab*-plane). Our theoretical analysis, using models for both two-band systems and a newly developed single-band incorporating vHs and gap anisotropy, reveals that the observed upper critical field behavior is driven by the anisotropy of the Fermi velocity originating from van Hove singularities rather than multi-band effects or gap anisotropy. Increased electron scattering via proton irradiation defects smears out the van Hove singularities and restores the conventional behavior. Our work underscores the critical role of 'singularities' in the density-of-states near the Fermi energy to the superconducting behavior.

## ACKNOWLEDGMENTS

This work was supported by the U. S. Department of Energy, Office of Science, Basic Energy Sciences, Materials Sciences and Engineering Division. The proton irradiation was performed at Western Michigan University.

## REFERENCES

1. D. E. Freedman, T. H. Han, A. Prodi, P. Muller, Q.- Z. Huang, Y.-S. Chen, S. M. Webb, Y. S. Lee, T. M. McQueen, and D. G. Nocera, Site specific x-ray anomalous dispersion of the geometrically frustrated kagome magnet, herbertsmithite, $ZnCu_3(OH)_6C_{l2}$, *J. Am. Chem. Soc*. **132**, 16185 (2010).
2. T.-H. Han, J. S. Helton, S. Chu, D. G. Nocera, J. A. Rodriguez-Rivera, C. Broholm, and Y. S. Lee, Fractionalized excitations in the spin-liquid state of a kagome-lattice antiferromagnet, *Nature* **492**, 406 (2012).
3. L. Ye, M. Kang, J. Liu, F. von Cube, C. R. Wicker, T. Suzuki, C. Jozwiak, A. Bostwick, E. Rotenberg, D. C. Bell, L. Fu, R. Comin, and J. G. Checkelsky, Massive Dirac fermions in a ferromagnetic Kagome metal, *Nature* **555**, 638 (2018).
4. S. V. Isakov, S. Wessel, R. G. Melko, K. Sengupta, and Y. B. Kim, Hard-Core Bosons on the Kagome Lattice: Valence-Bond Solids and Their Quantum Melting, *Phys. Rev. Lett*. **97**, 147202 (2006).
5. K. Nakayama, Y. Li, M. Liu, Z. Wang, T. Takahashi, Y. Yao, and T. Sato, Multiple energy scales and anisotropic energy gap in the charge-density-wave phase of the kagome superconductor $CsV_3Sb_5$, *Phys. Rev. B* **104**, L161112 (2021).
6. T. Neupert, M. M. Denner, J. X. Yin, R. Thomale, and M. Z. Hasan, Charge order and superconductivity in kagome materials, *Nat. Phys.* **18,** 137 (2022).
7. B. R. Ortiz, L. C. Gomes, J. R. Morey, M. Winiarski, M. Bordelon, J. S. Mangum, I. W. H. Oswald, J. A. Rodriguez-Rivera, J. R. Neilson, S. D. Wilson, E. Ertekin, T. M. McQueen, and E. S. Toberer, New kagome prototype materials: discovery of $KV_3Sb_5$, $RbV_3Sb_5$, and $CsV_3Sb_5$, *Phys. Rev. Mater*. **3**, 094407 (2019).
8. B. R. Ortiz, S. M. L. Teicher, Y. Hu, J. L. Zuo, P. M. Sarte, E. C. Schueller, A. M. Milinda Abeykoon, M. J. Krogstad, S. Rosenkranz, R. Osborn, R. Seshadri, L. Balents, J. He, and S. D. Wilson, $CsV_3Sb_5$: A $Z_2$ Topological Kagome metal with a superconducting ground state, *Phys. Rev. Lett*. **125**, 247002 (2020).
9. Y. Fu, N. Zhao, Z. Chen, Q. Yin, Z. Tu, C. Gong, C. Xi, X. Zhu, Y. Sun, K. Liu, and H. Lei, Quantum transport evidence of topological band structures of Kagome superconductor $CsV_3Sb_5$, *Phys. Rev. Lett.* **127**, 207002 (2021).
10. F. H. Yu, T. Wu, Z. Y. Wang, B. Lei, W. Z. Zhuo, J. J. Ying, and X. H. Chen, Concurrence of anomalous Hall effect and charge density wave in a superconducting topological kagome metal, *Phys. Rev. B* **104**, L041103 (2021).
11. S.-Y. Yang, Y. Wang, B. R. Ortiz, D. Liu, J. Gayles, E. Derunova, R. Gonzalez-Hernandez, L. Šmejkal, Y. Chen, S. S. P. Parkin, S. D. Wilson, E. S. Toberer, T. McQueen, M. N. Ali, Giant, unconventional anomalous Hall effect in the metallic frustrated magnet candidate, $KV_3Sb_5$, *Sci. Adv.* **6**, eabb6003 (2020).

12. Z. Liang, X. Hou, F. Zhang, W. Ma, P. Wu, Z. Zhang, F. Yu, J.-J. Ying, K. Jiang, L. Shan, Z. Wang, and X.-H. Chen, Three-dimensional charge density wave and surface-dependent vortex-core states in a kagome superconductor $CsV_3Sb_5$, *Phys. Rev. X* **11**, 031026 (2021).

13. F. H. Yu, D. H. Ma, W. Z. Zhuo, S. Q. Liu, X. K. Wen, B. Lei, J. J. Ying, and X. H. Chen, Unusual competition of superconductivity and charge-density-wave state in a compressed topological kagome metal, *Nat. Commun*. **12**, 3645 (2021).

14. Z. Y. Zhang, Z. Chen, Y. Zhou, Y. F. Yuan, S. Y. Wang, J. Wang, H. Y. Yang, C. An, L. L. Zhang, X. D. Zhu, Y. H. Zhou, X. L. Chen, J. H. Zhou, and Z. R. Yang, Pressure-induced reemergence of superconductivity in the topological kagome metal $CsV_3Sb_5$, *Phys. Rev. B* **103**, 224513 (2021).

15. J. Ge, P. Wang, Y. Xing, Q. Yin, A. Wang, J. Shen, H. Lei, Z. Wang, and J. Wang, Charge-4e and charge-6e flux quantization and higher charge superconductivity in kagome superconductor ring devices, *Phys. Rev. X.* **14**, 021025 (2024).

16. S. L. Ni, S. Ma, Y. H. Zhang, J. Yuan, H. T. Yang, Z. Y. W. Lu, N. N. Wang, J. P. Sun, Z. Zhao, D. Li *et al.*, Anisotropic superconducting properties of kagome metal $CsV_3Sb_5$, *Chin. Phys. Lett*. **38**, 057403 (2021).

17. W. Duan, Z. Nie, S. Luo, F. Yu, B. R. Ortiz, L. Yin, H. Su, F. Du, A. Wang, Y. Chen, X. Lu, J. Ying, S. D. Wilson, X. Chen, Y. Song, and H. Yuan, Nodeless superconductivity in the kagome metal $CsV_3Sb_5$, *Sci. China-Phys. Mech. Astron*. **64**, 107462 (2021).

18. H. D. Scammell, J. Ingham, T. Li, and O. P. Sushkov, Chiral excitonic order from twofold van Hove singularities in kagome metals, *Nat. Commun*. **14**, 605 (2023).

19. H. Tan, Y. Li, Y. Liu, D. Kaplan, Z. Wang, and B. Yan, Emergent topological quantum orbits in the charge density wave phase of kagome metal $CsV_3Sb_5$, *npj Quantum Materials* **8**, 39 (2023).

20. B. R. Ortiz, S. M. L. Teicher, L. Kautzsch, P. M. Sarte, J. P. C. Ruff, R. Seshadri, and S. D. Wilson, Fermi surface mapping and the nature of charge-density-wave order in the Kagome superconductor $CsV_3Sb_5$, *Phys. Rev. X* **11**, 041030 (2021).

21. M. Kang, S. Fang, J.-K. Kim, B. R. Ortiz, S. H. Ryu, J. Kim, J. Yoo, G. Sangiovanni, D. Di Sante et al, Twofold van Hove singularity and origin of charge order in topological kagome superconductor $CsV_3Sb_5$, *Nat. Phys*. **18**, 301 (2022).

22. Y. Hu, X. Wu, B. R. Ortiz, S. Ju, X. Han, J. Ma, N. C. Plumb, M. Radovic, R. Thomale, S. D. Wilson, A. P. Schnyder, and M. Shi, Rich nature of van Hove singularities in kagome superconductor $CsV_3Sb_5$, *Nat. Commun*. **13**, 2220 (2022).

23. H. LaBollita and A. S. Botana, Tuning the van Hove singularities in $AV_3Sb_5$ (A = K, Rb, Cs) via pressure and doping, *Phys. Rev. B* **104**, 205129 (2021).

24. Y. Luo, Y. Han, J. Liu, H. Chen, Z. Huang, L. Huai, H. Li, B. Wang, J. Shen, S. Ding et al, A unique van Hove singularity in kagome superconductor $CsV_{3-x}Ta_xSb_5$ with enhanced superconductivity, *Nat. Commun*. **14**, 3819 (2023).

25. T. Kato, Y. Li, T. Kawakami, M. Liu, K. Nakayama, Z. Wang, A. Moriya, K. Tanaka, T. Takahashi, Y. Yao, T. Sato, Three-dimensional energy gap and origin of charge-density wave in Kagome superconductor $KV_3Sb_5$, *Commun. Mater*. **3**, 30 (2022).

26. Y. Luo, S. Peng, S. M. L. Teicher, L. Huai, Y. Hu, Y. Han, B. R. Ortiz, Z. Liang, Z. Wei, J. Shen *et al.,* Electronic states dressed by an out-of-plane supermodulation in the quasi-two-dimensional kagome superconductor $CsV_3Sb_5$, *Phys. Rev. B* **105**, L241111 (2022).

27. S. D. Wilson and B. R. Ortiz, $AV_3Sb_5$ kagome superconductors, *Nat. Rev. Mater*. **9**, 420 (2024).

28. C. Wang, S. Liu, H. Jeon and J.-H. Cho, Origin of charge density wave in the layered kagome metal CsV$_3$Sb$_5$, *Phys. Rev. B* **105**, 045135 (2022).

29. W.-S. Wang, Z.-Z. Li, Y.-Y. Xiang and Q. -H. Wang, Competing electronic orders on kagome at van Hove filling, *Phys. Rev. B* **87**, 115135 (2013).

30. T. Nguyen and M. Li, Electronic properties of correlated kagome metals AV$_3$Sb$_5$ (A = K, Rb, Cs): A perspective, *J. Appl. Phys*. **131**, 060901 (2022).

31. A. E. Koshelev, R. Chapai, D. Y. Chung, J. F. Mitchell, and U. Welp, Origin of anomalous magnetotransport in kagome superconductors AV$_3$Sb$_5$ (A = K, Rb, Cs), *Phys. Rev. B* **110**, 024512 (2024).

32. V. G. Kogan, R. Prozorov, and A. E. Koshelev, Temperature-dependent anisotropies of upper critical field and London penetration depth, *Phys. Rev. B* **100**, 014518 (2019).

33. M. Zehetmayer, A review of two-band superconductivity: materials and effects on the thermodynamic and reversible mixed-state properties, *Supercond. Sci. Technol*. **26**, 043001 (2013).

34. H.-S. Xu, Y.-J. Yan, R. Yin, W. Xia, S. Fang, Z. Chen, Y. Li, W. Yang, Y. Guo, and D. -L. Feng, Multiband Superconductivity with sign-preserving order parameter in Kagome superconductor CsV$_3$Sb$_5$, *Phys. Rev. Lett*. **127**, 187004 (2021).

35. J. Liu, Q. Li, Y. Li, X. Fan, J. Li, P. Zhu, H. Deng, J.-X. Yin, H. Yang, J. Li, and H. -H. Wen, Enhancement of superconductivity and phase diagram of Ta-doped Kagome superconductor CsV$_3$Sb$_5$, *Sci. Rep*. **14**, 9580. (2024).

36. Y. Zhong, J. Liu, X. Wu, Z. Guguchia, J.-X. Yin, A. Mine, Y. Li, S. Najafzadeh, D. Das, C. Mielke III *et al*., Nodeless electron pairing in CsV$_3$Sb$_5$-derived kagome superconductors, *Nature* **617**, 488 (2023).

37. A. Mine, Y. Zhong, J. Liu, T. Suzuki, S. Najafzadeh, T. Uchiyama, J.-X. Yin, X. Wu, X. Shi, Z. Wang,Y. Yao, and K. Okazaki, Direct observation of orbital-selective anisotropic Cooper pairing in kagome superconductor CsV$_3$Sb$_5$, arXiv:2404.18472.

38. M. J. Grant, Y. Liu, G.-H. Cao, J. A. Wilcox, Y. Guo, X. Xu, and A. Carrington, Superconducting energy gap structure of CsV$_3$Sb$_5$ from magnetic penetration depth measurements, arXiv:2411.05611 (cond-mat) (2024), https://arxiv.org/abs/2411.05611

39. R. Gupta, D. Das, C. H. Mielke III, Z. Guguchia, T. Shiroka, C. Baines, M. Bartkowiak, H. Luetkens, R. Khasanov, Q. Yin, Z. Tu, C. Gong, and H. Lei, Microscopic evidence for anisotropic multigap superconductivity in the CsV$_3$Sb$_5$ Kagome superconductor, *npj Quantum materials* **7**, 49 (2022).

40. Z. Shan, P. K. Biswas, S. K. Ghosh, T. Tula, A. D. Hillier et al., Muon spin relaxation study of the layered Kagome superconductor CsV$_3$Sb$_5$, *Phys. Rev. Res*. **4**, 033145 (2022).

41. L. R. Testardi, L. F. Mattheiss, Electron lifetime effects on properties of A15 and bcc materials, *Phys. Rev. Lett*. **41**, 1612 (1978).

42. L. F. Mattheiss, L. R. Testardi, Electron-lifetime effects on properties of Nb3Sn, Nb3Ge, and Ti-V-Cr alloys, *Phys. Rev. B* **20**, 2196 (1979).

43. M. Putti, R. Vaglio, J. M. Rowell, Radiation effects on MgB$_2$: a review and a comparison with A15 superconductors, *Supercond. Sci. Technol.* **21**, 043001 (2008).

44. N. R. Werthamer, E. Helfand, and P. C. Hohenberg, Temperature and purity dependence of the superconducting critical field, $H_{c2}$. III. Electron spin and spin-orbit Effects, *Phys. Rev*. **147**, 295 (1966).

45. K. Maki, The magnetic properties of superconducting alloys I, *Phys*. **1**, 21 (1964).

46. P. G. de Gennes, Behavior of dirty superconductors in high magnetic fields, *Phys. Kondens. Materie.* **3**, 79 (1964).

47. See Supplemental Material for basic characterization, temperature dependence of magnetization and TDO frequency, for both pristine and irradiated samples as various proton doses, angular dependence of in-plane $H_{c2}(T)$, and additional details on upper critical field models for two-gap system and for open Fermi surface close to van Hove singularities.

48. R. Chapai, M. Leroux, V. Oliviero, D. Vignolles, N. Bruyant, M. P. Smylie, D. Y. Chung, M. G. Kanatzidis, W.-K. Kwok, J. F. Mitchell, U. Welp, Magnetic breakdown and topology in the Kagome superconductor $CsV_3Sb_5$ under high magnetic field, *Phys. Rev. Lett.* **130**, 126401 (2023).

49. K. Shrestha, R. Chapai, B. K. Pokharel, D. Miertschin, T. Nguyen *et al.*, Nontrivial Fermi surface topology of the Kagome superconductor $CsV_3Sb_5$ probed by de Haas-van Alphen oscillations, *Phys. Rev. B* **105**, 024508 (2022).

50. M. P. Smylie, K. Willa, H. Claus, A. Snezhko, I. Martin, W.-K. Kwok, Y. Qiu, Y. S. Hor, E. Bokari, P. Niraula, A. Kayani, V. Mishra, and U. Welp, Robust odd-parity superconductivity in the doped topological insulator $Nb_xBi_2Se_3$, *Phys. Rev. B* **96**, 115145 (2017).

51. Y. Wu, Q. Wang, X. Zhou, J. Wang, P. Dong, J. He, Y. Ding, B. Teng, Y. Zhang, Y. Li, *et al.*, Nonreciprocal charge transport in topological kagome superconductor $CsV_3Sb_5$, *npj Quantum Materials* **7**, 105 (2022).

52. X. Huang, C. Guo, C. Putzke, M. Gutierrez-Amigo, Y. Sun, M. G. Vergniory, I. Errea, D. Chen, C. Felser and P. J. Moll, Three-dimensional Fermi surfaces from charge order in layered $CsV_3Sb_5$, *Phys. Rev. B* **106**, 064510 (2022).

53. Y. Xiang, Q. Li, Y. Li, W. Xie, H. Yang, Z. Wang, Y. Yao, and H. -H. Wen, Twofold symmetry of $c$-axis resistivity in topological kagome superconductor $CsV_3Sb_5$ with in-plane rotating magnetic field, *Nat. Commun.* **12**, 6727 (2021).

54. Z. Wang, Y.-X. Jiang, J.-X. Yin, Y. Li, G.-Y. Wang, H.-L. Huang, S. Shao, J. Liu, P. Zhu, N. Shumiya, *et al.*, Electronic nature of chiral charge order in the kagome superconductor $CsV_3Sb_5$, *Phys. Rev. B* **104**, 075148 (2021).

55. J. Li, Q. Li, Y. Xiang, H. Yang, Z. Wang, Y. Yao, and H.-H. Wen, Local pairing versus bulk superconductivity intertwined by the charge density wave order in $Cs(V_{1-x}Ta_x)_3Sb_5$, *Phys. Rev. Mater.* **8**, 014801 (2024).

56. A. Gurevich, Enhancement of the upper critical field by nonmagnetic impurities in dirty two-gap superconductors, *Phys. Rev. B* **67**, 184515 (2003).

57. K. H. Müller, G. Fuchs, A. Handstein, K. Nenkov, V. N. Narozhnyi and D. Eckert, The upper critical field in superconducting $MgB_2$, *J. Alloys Compd*, **322**, L10 (2001).

58. L. Lyard, P. Samuely, P. Szabo, T. Klein, C. Marcenat, L. Paulius, K. H. P. Kim, C. U. Jung, H.-S. Lee, B. Kang, *et al.*, Anisotropy of the upper critical field and critical current in single crystal $MgB_2$, *Phys. Rev. B* **66**, 180502 (2002).

59. A. A. Golubov, and A. E. Koshelev, Upper critical field in dirty two-band superconductors: Breakdown of the anisotropic Ginzburg-Landau theory, *Phys. Rev. B* **68**, 104503 (2003).

60. D. W. Younger, and R. A. Klemm, Theory of the upper critical field in anisotropic superconductors, *Phys. Rev.* **21**, 3890 (1980).

61. P. C. Hohenberg, and N. R. Werthamer, Anisotropy and temperature dependence of the upper critical field of type-II superconductors, *Phys. Rev.* **153**, 493 (1967).

62. H. Teichler, On the theory of $H_{c2}$ anisotropy in cubic superconductors, *Phys. Stat. Sol. (b)* **69**, 501 (1975).
63. P. Miranovic, K. Machida, V. G. Kogan, Anisotropy of the upper critical field in superconductors with anisotropic gaps: Anisotropy parameters of $MgB_2$, *J. Phys. Soc. Jpn.* **72**, 221 (2003).
64. J. Wang, X. Xu, N. Zhou, L. Li, X. Cao, J. Yang, Y. Li, C. Cao, J. Dai, J. Zhang, Z. Shi, B. Chen, and Z. Yang, Upward curvature of the upper critical field and the V-shaped pressure dependence of Tc in the nocentrosymmetric superconductor $PbTaSe_2$, *J. Supercond. Nov. Magn.* **28**, 3173 (2015).
65. M. Bristow, A. Gower, J. C. A. Prentice, M. D. Watson, Z. Zajicek, S. J. Blundell, A. A. Haghighirad, A. McCollam, and A. I. Coldea, Multiband description of the upper critical field of bulk FeSe, *Phys. Rev.* B **108**, 184507 (2023).
66. S. Khim, B. Lee, J. W. Kim, E. S. Choi, G. R. Stewart, K. H. Kim, Pauli-limiting effects in the upper critical fields of a clean LiFeAs single crystal, *Phys. Rev. B* **84**, 104502 (2011).
67. M. P. Smylie, A. E. Koshelev, K. Willa, R. Willa, W.-K. Kwok, J.-K. Bao, D.-Y. Chung, M. G. Kanatzidis, J. Singleton, F. F. Balakirev *et al.*, Anisotropic upper critical field of pristine and proton-irradiated single crystals of the magnetically ordered superconductor $RbEuFe_4As_4$, *Phys. Rev. B* **100**, 054507 (2019).
68. M. Bristow, W. Knafo, P. Reiss, W. Meier, P. C. Canfield, S. J. Blundell, and A. I. Coldea, Competing pairing interactions responsible for the large upper critical field in a stoichiometric iron-based superconductor $CaKFe_4As_4$, *Phys. Rev. B* **101**, 134502 (2020).
69. V. G. Kogan, and R. Prozorov, Orbital upper critical field and its anisotropy of clean one-and two-band superconductors, *Rep. Prog. Phys.* **75**, 114502 (2012).
70. H. Q. Yuan, J. Singleton, F. F. Balakirev, S. A. Baily, G. F. Chen, J. L. Luo, and N. L. Wang, Nearly isotropic superconductivity in $(Ba,K)Fe_2As_2$, *Nature* **457**, 565 (2009).
71. K.-H. Müller, G. Fuchs, A. Handstein, K. Nenkov, V. N. Narozhnyi, and D. Eckert, The upper critical field in superconducting $MgB_2$, *J. Alloys Compd* **322,** L11 (2001).
72. K. Fukushima, K. Obata, S. Yamane, Y. Hu, Y. Li, Y. Yao, Z. Wang, Y. Maeno, S. Yonezawa, Violation of emergent rotational symmetry in the hexagonal Kagome superconductor $CsV_3Sb_5$, *Nat. Commun.* **15**, 2888 (2024).
73. M. S. Hossain, Q. Zhang, E. S. Choi, D. Ratkovski, B. Lüscher, Y. Li, Y.-X. Jiang, M. Litskevich, Z.-J. Cheng, J.-X. Yin, T. A. Cochran, B. Casa, B. Kim, X. Yang, J. Liu, Y. Yao, A. F. Bangura, Z. Wang, M. H. Fischer, T. Neupert, L. Balicas, M. Z. Hasan, Unconventional gapping behavior in a Kagome superconductor, *Nat. Phys.* **21**, 556 (2025).
74. R. G. Dias and J. M. Wheatley, Superconducting upper critical field near a 2D van Hove singularity, *Solid State Commun.*, **98**, 859 (1996).
75. R. O. Zaitsev, On the effect of van Hove singularities on the critical field of type-II superconductors, *JETP Letters*, **65**, 74 (1997). (Pis'ma Zh. Eksp. Teor. Fiz. **65**, 71 (1997)).
76. R. G. Dias, Effects of van Hove singularities on the upper critical field, *J. Phys.: Condens. Matter.* **12**, 9053 (2000).
77. M. Roppongi, K. Ishihara, Y. Tanaka, K. Ogawa, K. Okada, S. Liu, K. Mukasa, Y. Mizukami, Y. Uwatoko, R. Grasset, M. Konczykowski, B. R. Ortiz, S. D. Wilson, K. Hashimoto, and T. Shibauchi, Bulk evidence of anisotropic s-wave pairing with no sign change in the kagome superconductor $CsV_3Sb_5$, *Nat. Commun.* **14**, 667 (2023).
78. H. J. Choi, D. Roundy, H. Sun, M. L. Cohen, S. G. Louie, The origin of the anomalous superconducting properties of $MgB_2$, *Nature* **418**, 758 (2002).

79. C. Buzea, T. Yamashita, Review of the superconducting properties of MgB$_2$, *Supercond. Sci. Technol.* **14**, R115 (2001).

80. Y. Li, Q. Li, X. Fan, J. Liu, Q. Feng, M. Liu, C. Wang, J.X. Yin, J. Duan, X. Li, Z. Wang, H.-H. Wen, Y. Yao, Tuning the competition between superconductivity and charge order in the Kagome superconductor Cs(V$_{1-x}$Nb$_x$)$_3$Sb$_5$, *Phys. Rev. B* **105**, L180507 (2022).

81. J. Liu, Q. Li, Y. Li, X. Fan, J. Li, P. Zhu, H. Deng, J.-X. Yin, H. Yang, H.-H. Wen, Z. Wang, Enhancement of superconductivity and phase diagram of Ta-doped Kagome superconductor CsV$_3$Sb$_5$, *Sci. Rep.* **14**, 9580 (2024).

82. Z. Zhao, R. Wang, Y. Zhang, K. Zhu, W. Yu, Y. Han, J. Liu, G. Hu, H. Guo, X. Lin, X. Dong, Hui Chen, H. Yang, H.-J. Gao, Two-fold symmetry of the in-plane resistance in Kagome superconductor Cs(V$_{1-x}$Ta$_x$)$_3$Sb$_5$ with enhanced superconductivity, *Chin. Phys. B* **33**, 077406 (2024).

83. T. Kato, Y. Li, K. Nakayama, Z. Wang, S. Souma, F. Matsui, M. Kitamura, K. Horiba, H. Kumigashira, T. Takahashi, Y. Yao, T. Sato, Fermiology and origin of $T_c$ enhancement in a Kagome superconductor Cs(V$_{1-x}$Nb$_x$)$_3$Sb$_5$, *Phys. Rev. Lett.* **129**, 206402 (2022).

84. Y. Luo, Y. Han, J. Liu, H. Chen, Z. Huang, L. Huai, H. Li, B. Wang, J. Shen, S. Ding, Z. Li, S. Peng, Z. Wei, Y. Miao, X. Sun, Z. Ou1, Z. Xiang, M. Hashimoto, D. Lu, Y. Yao, H. Yang, X. Chen, H.-J. Gao, Z. Qiao, Z. Wang, J. He, A unique van Hove singularity in Kagome superconductor CsV$_{3-x}$Ta$_x$Sb$_5$ with enhanced superconductivity, *Nat. Commun.* **14**, 3819 (2023).

85. M. Leroux, V. Mishra, J. P. C. Ruff, H. Claus, M. P. Smylie, C. Opagiste, P. Rodiere, A. Kayani, G. D. Gu, J. M. Tranquada, W. K. Kwok, Z. Islam, and U. Welp, Disorder raises the critical temperature of a cuprate superconductor, *Proc. Natl. Acad. Sci. USA* **116**, 10691 (2019).

86. F. Rullier-Albenque, H. Alloul, and R. Tourbot, Influence of pair breaking and phase fluctuations on disordered high $T_c$ cuprate superconductors, *Phys. Rev. Lett.* **91**, 047001 (2003).

87. Y. Feng, J. Wang, R. Jaramillo, J. v. Wezel, S. Haravifard *et al.*, Order parameter fluctuations at a buried quantum critical point, *Proc. Natl. Acad. Sci. USA* **109**, 7224 (2012).

88. B. Sipos, A. F. Kusmartseva, A. Akrap, H. Berger, L. Forro, E. Tutis, From Mott state to superconductivity in 1T-TaS$_2$, *Nat. Mater.* **7**, 960 (2008).

89. K. Y. Chen, N. N. Wang, Q. W. Yin, Y. H. Gu, K. Jiang, Z. J. Tu, C. S. Gong, Y. Uwatoko, J. P. Sun, H. C. Lei, J. P. Hu, and J.-G. Cheng, Double superconducting dome and triple enhancement of $T_c$ in the Kagome superconductor CsV$_3$Sb$_5$ under high pressure, *Phys. Rev. Lett.* **126**, 247001 (2021).

90. E. I. Timmons, S. Teknowijoyo, M. Konczykowski, O. Cavani, M. A. Tanatar, S. Ghimire, K. Cho, Y. Lee, L. Ke, N. H. Jo, S. L. Bud'ko, P. C. Canfield, P. P. Orth, M. S. Scheurer, and R. Prozorov, Electron irradiation effects on superconductivity in PdTe$_2$: An application of a generalized Anderson theorem, *Phys. Rev. Res.* **2**, 023140 (2020).

**FIGURES**

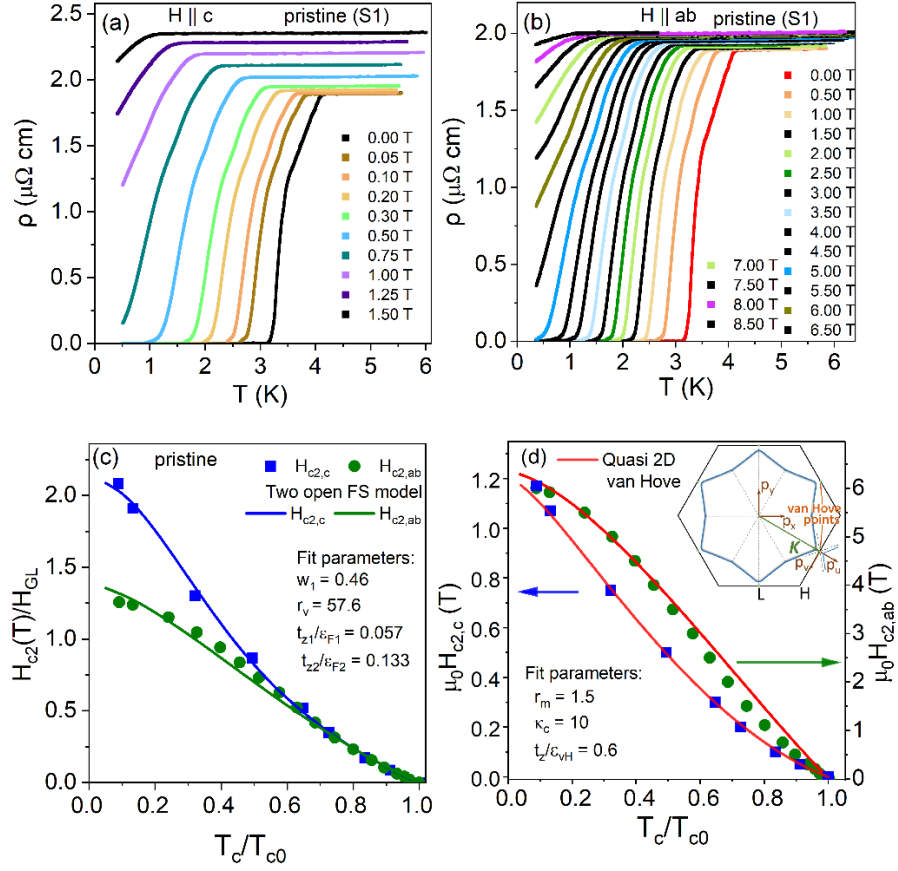

FIG. 1. Temperature dependence of the in-plane electrical resistivity $\rho(T)$ of pristine $CsV_3Sb_5$ (sample S1) through the superconducting transition (a) with $H \parallel c$ and (b) $H \parallel ab$ at various fields as indicated. (c) Superconducting phase diagram with field applied along $c$-axis (blue squares) and $ab$-plane (green dots), both normalized with their respective GL values of $H_{c2,c}^{GL} = 0.6$ T and $H_{c2,ab}^{GL} = 4.8$ T. The solid lines represent fits based on the model of two open Fermi surfaces (Eq. (1)). (d) $H_{c2}$ vs $T_c/T_{c0}$ for both orientations, fitted with the quasi-2D van Hove model (Eq. (2)). *Inset*: Schematic cross-section of the Fermi surface used in the vHs model, Eq. (2), for a given wave vector along the $c$-axis as described in detail in the Supplemental Materials [47]. Here, $p_u$ and $p_v$ are the components of the electron momentum along and across the vHs.

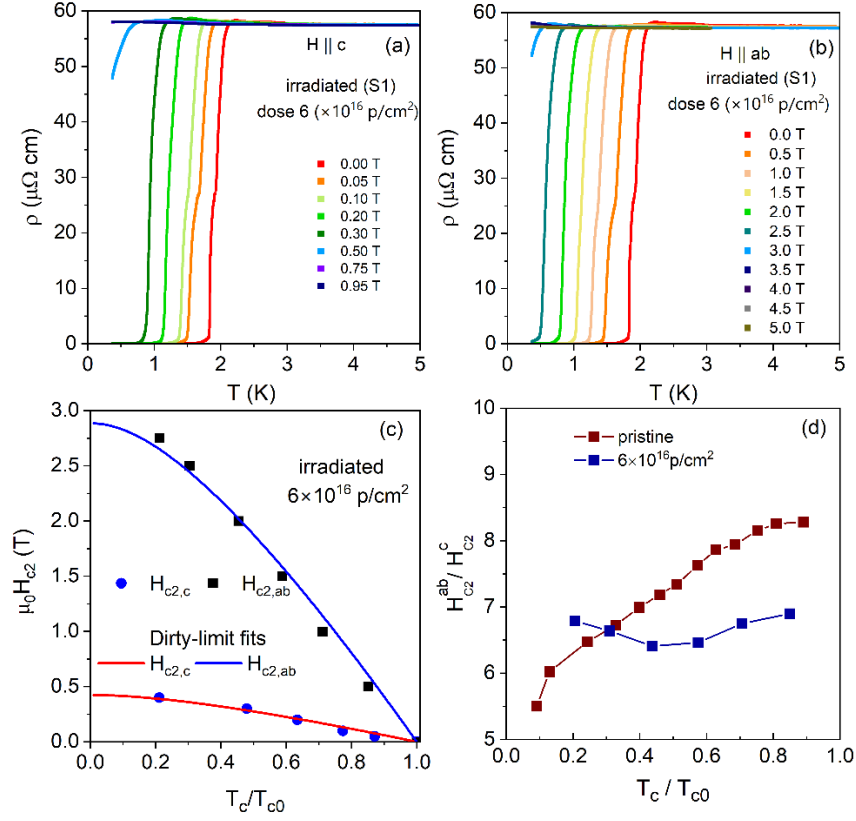

FIG. 2. The $\rho(T)$ of a crystal of CsV$_3$Sb$_5$ (sample S1) through the superconducting transition (a) with $H \parallel c$ and (b) with $H \parallel ab$ following an irradiation dose of $6 \times 10^{16}$ p/cm$^2$. (c) $H_{c2}$ vs $T$ obtained from the data in frames (a-b). Solid lines are dirty-limit fits [41]. (d) Temperature-dependent anisotropy of the pristine (red) and temperature-independent anisotropy of the irradiated sample (blue).

**END MATTER**

*Appendix: Effect of consecutive irradiation runs*

The resistivity data described above were supplemented with magnetization and Tunnel Diode Oscillator (TDO) measurements on three crystals (S2, S3, S4) that underwent a series of consecutive irradiation runs. In the following, we highlight the observed phenomena with selected data; the complete data sets can be found in the Supplemental Materials [47]. The temperature dependence of magnetization $M(T)$ of sample S2 for various cumulative proton doses is shown in Fig. 3(a). A clear suppression of the superconducting transition is observed with increasing proton dose. Like $T_c$, the signature of $T_{CDW}$ gradually weakens and eventually becomes unobservable above around $\approx 4\times10^{16}$ p/cm$^2$, (Fig. 3(b)).

Figure 3(c) displays the variation of $T_c$ and $T_{CDW}$, normalized to their values in the pristine state, plotted as a function of proton dose. With increasing dose, both the superconducting and CDW transitions are suppressed in a nonlinear fashion, with the rate of suppression decreasing with increasing dose, such that at higher doses $T_c$ appears to approach saturation. The simultaneous suppression of CDW and reduction of the $T_c$, as observed in Fig. 3(c), is unexpected for a scenario where CDW order and superconductivity compete for the same electronic density of states at the Fermi level, as seen, for instance, in cuprate and dichalcogenide superconductors [85-88]. Given that $T_c$ and $T_{CDW}$ are well separated in CsV$_3$Sb$_5$, it is more likely that the observed reduction in both $T_c$ and $T_{CDW}$ is caused by increased scattering, rather than a direct competition between the two orders [13, 14, 89].

Figure 3(d) presents the $H_{c2}(T)$ dependences as extracted from the TDO data extending the susceptibility measurements down to 0.4 K. Consistent with our magnetization measurements (Fig. 3(a)), the transition temperature as observed via TDO measurement decreases with increasing irradiation dose. Measurements in various applied magnetic fields (see Fig. S5 [47]), the phase boundary $H_{c2,c}(T)$ vs $T$ is constructed for the pristine and for irradiated samples (for $H \parallel c$). The upward curvature in the $H_{c2}(T)$ gradually disappears with increasing proton dose (also see Fig. S5(g) [47] for $H \parallel ab$), transitioning to the conventional behavior and consistent with our model of a progressive smearing of the van Hove singularities.

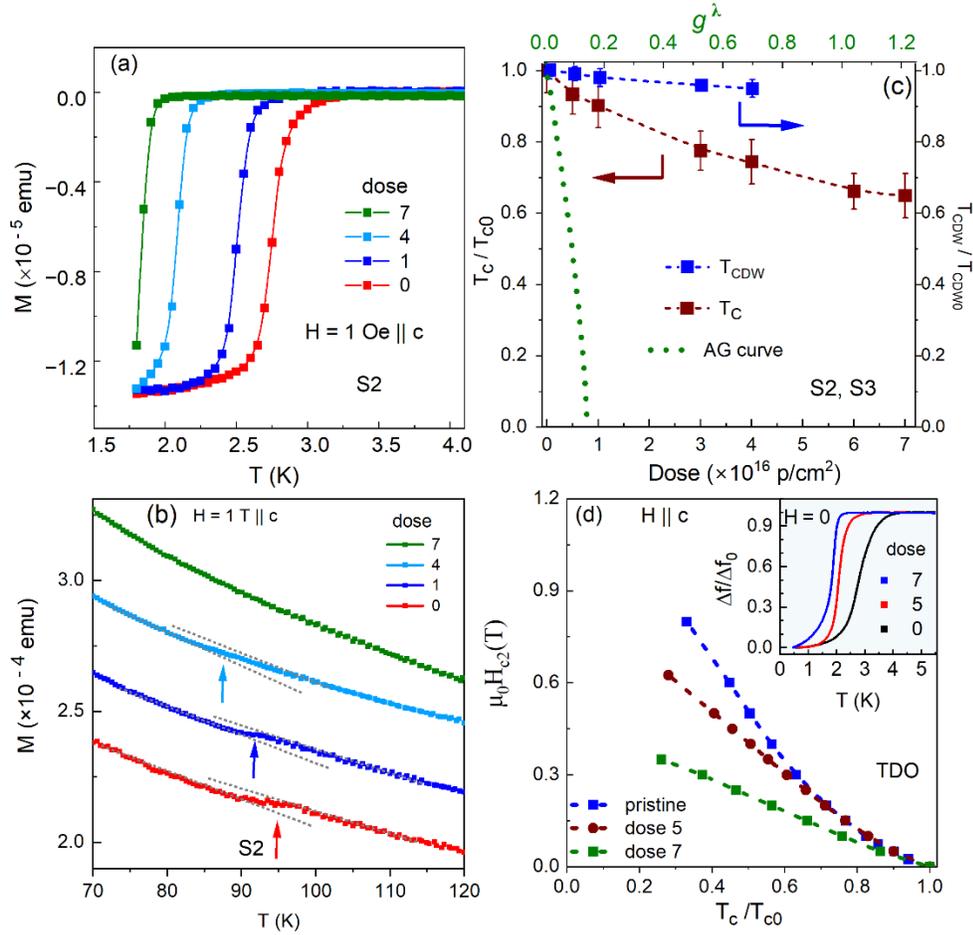

FIG. 3. (a) The superconducting transition of sample S2 as seen in the temperature dependence of the magnetic moment $M(T)$ for various proton irradiation doses. (b) $M(T)$ between 70 K and 120 K (through the CDW transition marked by the arrows). (c) Transition temperatures $T_{CDW}$ and $T_c$ (both normalized to the values of the pristine sample) of samples S2 and S3 vs dose. The error bars represent the width of the transition. The dashed line (green) represents the suppression of $T_c$ as function of the dimensionless scattering parameter $g^\lambda$ (top x-axis) as expected in the Abrikosov-Gorkov theory for pair-breaking scattering [90] and discussed in more detail in the Supplemental Materials section [47]. (d) $H_{c2}(T)$ obtained from the TDO measurements for different dose levels for $H \parallel c$ showing the progressive transformation of the upwards curvature of $H_{c2}$ into the conventional linear variation. Inset: Normalized TDO frequency shift vs temperature at multiple irradiation doses. The numbers specifying the irradiation dose are in units of $10^{16}$ p/cm$^2$.

# Supplemental Materials

## Origin of the Unusual Temperature Dependence of the Upper Critical Field of Kagome Superconductor CsV$_3$Sb$_5$: Multiple Bands or van Hove Singularities?

Ramakanta Chapai,[1,] A. E. Koshelev,[2,] M. P. Smylie,[1,3] D. Y. Chung,[1] Asghar Kayani,[4] Khushi Bhatt,[5] Gaurab Rimal,[4] M. G. Kanatzidis,[1,6] W.-K. Kwok,[1] J. F. Mitchell,[1] and Ulrich Welp[1]

[1]*Materials Science Division, Argonne National Laboratory, Lemont, IL 60439, USA*
[2]*Department of Physics and Astronomy, University of Notre Dame, Notre Dame, IN 46656, USA*
[3]*Department of Physics and Astronomy, Hofstra University, Hempstead, NY 11549, USA*
[4]*Department of Physics, Western Michigan University, Kalamazoo, MI 49008, USA*
[5]*Physics Division, Argonne National Laboratory, Lemont, IL 60439, USA*
[6]*Department of Chemistry, Northwestern University, Evanston, IL 60201, USA*

**Basic characterization**

Electrical resistivity measurements of single crystalline CsV$_3$Sb$_5$ were performed in a 9-1-1 T triple-axis vector magnet inside a dilution refrigerator (Bluefors LD400) using the standard four-probe method with current applied along the *ab*-plane direction. The temperature dependence of the electrical resistivity, ρ(*T*), displays two anomalies: one at $T_c \approx$ 3.5 K, corresponding to superconductivity, and another at $T_{CDW} \approx$ 94 K associated with the CDW transition (see FIG. S1(a)). These anomalies are also clearly seen in the temperature dependence of the magnetization *M*(*T*) (shown in Fig. S1(b)) and are consistent with previous studies [1-3]. Moreover, the low residual resistivity ($\rho_0 \sim$ 1.2 μΩ cm at 5 K) and high residual resistivity ratio ($RRR \approx$ 52) in ρ(*T*) reflect the high quality of our single crystal specimens. To obtain the superconducting phase diagram, ρ(*T*) data (in Fig. 1(a-b) in the main text) were supplemented with field dependent measurements, ρ(*H*) (Fig.S1(c, d)), at base temperature ~ 28 mK.

Tunnel diode oscillator (TDO) measurements were performed in a custom-built TDO system operating at ~14.5 MHz in a $^3$He cryostat. Irradiation of the samples with 5-MeV protons was performed at the tandem van de Graaff accelerator at Western Michigan University. A gold foil was used to disperse the beam to ensure a uniform beam spot over the samples, and the irradiation stage was cooled to ≈ -20 °C during irradiation. For the irradiation experiment, thin samples with typical size 0.5×0.5×0.05 mm$^3$ were selected. The Stopping and Range of Ions in Matter (SRIM) calculations [4] for our irradiation geometry show that proton implantation in the sample is negligible. One crystal (S1) was selected for detailed magneto-transport characterization in the pristine state and after irradiation to a relatively high dose of 6×10$^{16}$ p/cm$^2$.

The temperature dependence of the resistivity $\rho(T)$ of sample S1 following irradiation to a dose of $6 \times 10^{16}$ p/cm$^2$ is shown in Fig. S1(e). Upon irradiation, several features emerge: (i) the residual resistivity increases significantly, from $\rho_0 \approx 1.9$ μΩ cm (at 5 K) for pristine to $\rho_0 \approx 58$ μΩ cm for the irradiated sample. (ii) $T_c$ drops from 3.5 K (pristine) to 2.0 K (irradiated). (iii) The anomaly associated with the CDW transition becomes unobservable.

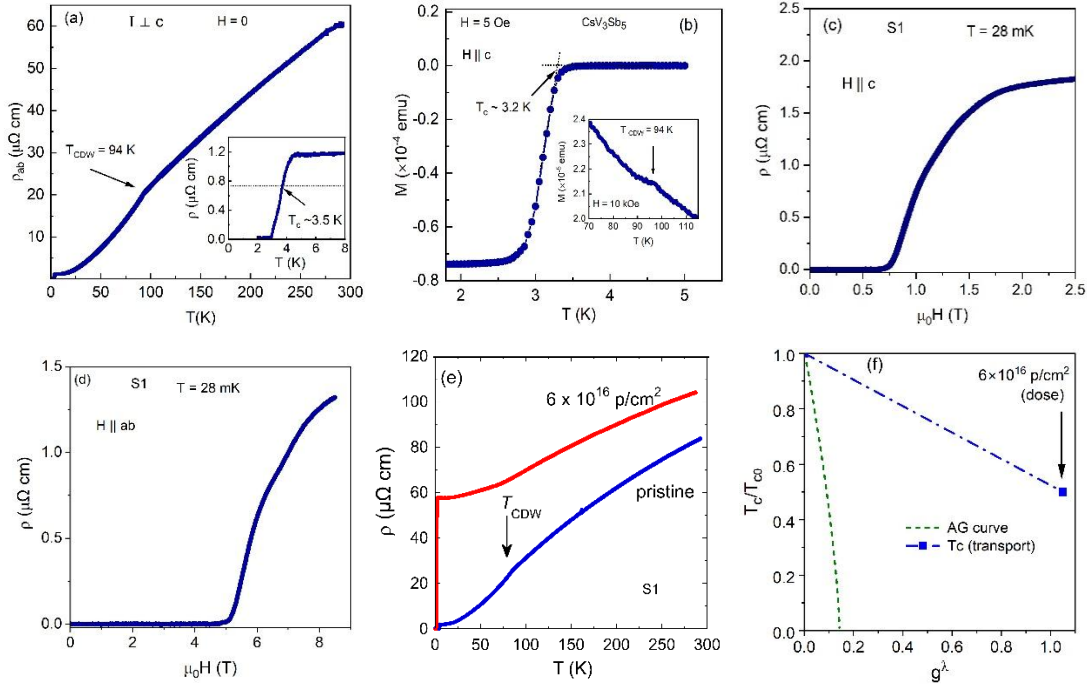

FIG. S1. (a) Temperature dependence of the in-plane electrical resistivity $\rho_{ab}(T)$ of pristine CsV$_3$Sb$_5$ between 300 K and 1.8 K displaying the CDW transition at $T_{CDW} \approx 94$ K. The inset shows the superconducting transition at $T_c \approx 3.5$ K. (b) Temperature dependence of the magnetization $M(T)$ between 1.8 K and 5 K measured by applying $H = 5$ Oe along the $c$ axis. Inset: $M(T)$ between 70 K and 115 K measured by applying $H = 10$ kOe along the $c$-axis and displaying the CDW transition around $T_{CDW} \approx 94$ K. Field dependence of the in-plane electrical resistivity $\rho_{ab}(T)$ of pristine CsV$_3$Sb$_5$ under (c) $H \parallel c$, and (d) $H \parallel ab$. (e) Temperature dependence of the in-plane electrical resistivity, $\rho_{ab}(T)$, of CsV$_3$Sb$_5$ sample S1; pristine (blue) and irradiated (red). At a dose of $6 \times 10^{16}$ p/cm$^2$, the CDW transition become unobservable. (f) Superconducting transition temperature (normalized to the values of the pristine sample) vs dimensionless scattering parameter ($g^\lambda$). Dashed line (green) represents the suppression of $T_c$ expected in the AG theory [5-7].

**Temperature dependence of the electrical resistivity, second crystal**

Figure S2 (a-b) presents the temperature-and field-dependent resistivity data measured on a second single crystal (sample S5), which yields identical results to those of sample S1(shown in Fig. 1, main text), indicating that the observed behavior is intrinsic. Specifically, for both orientations, the $H_{c2}(T)$ phase boundary displays pronounced upward curvature (see Fig. 1(c, d), main text and Fig. S2(c)). Note that while choosing a different criterion for $T_c$, will change the absolute values of $H_{c2}$, the shape of the $H_{c2}(T)$ phase boundary remains unchanged, as the resistive transitions uniformly shift with the application of a magnetic field (see Fig. S2 (c)) Also, both temperature dependent and field dependent data yield the same phase boundaries.

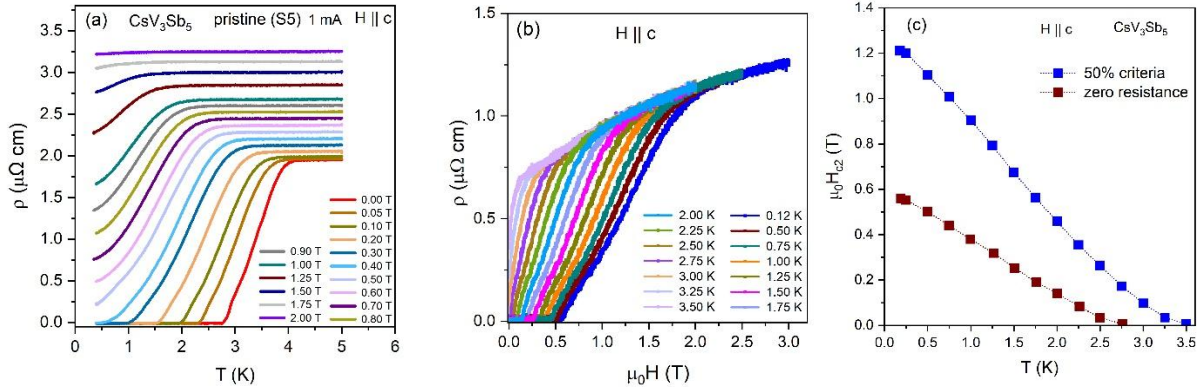

FIG. S2. (a) Temperature dependence of the electrical resistivity $\rho(T)$ of pristine $CsV_3Sb_5$ (sample S5) through the superconducting transition at various fields. Some structure is visible in the transition at low resistance values, well below the 50% mark. Nevertheless, the $H_{c2}$-data obtained on this sample are essentially identical to those on sample S1 shown in Fig. 1 of the main text, suggesting that the reported features are an intrinsic property of $CsV_3Sb_5$. (b) Field dependence of the electrical resistivity $\rho(H)$ of pristine $CsV_3Sb_5$ (sample S5) through the superconducting transition at various temperatures. (c) Superconducting phase diagram with field applied along $c$-axis. The phase diagram is constructed using two criteria, one with 50% resistance drop (blue) and other when zero resistance is reached (red).

**Temperature dependent magnetization, $M(T)$, of pristine and irradiated $CsV_3Sb_5$**

The temperature dependence of the magnetic moment $M(T)$ was measured in a superconducting quantum interference device (SQUID) magnetometer (MPMS3, Quantum Design) at various fields along the $c$-axis as presented in Figs. S3(a-d). The transition temperature $T_c$ is reduced upon the application of field, consistent with a superconducting transition. Using the $T_c$ values obtained from the onset of the transition in $M(T)$ at each field, the phase diagram of $H_{C2}(T)$ vs $T$ is constructed, as shown in Fig. S3(e). Note that, $T_c$ is suppressed rather abruptly until 100 Oe, and an upward curvature develops after above $T/T_c \sim 0.80$. Such characteristic features in phase diagrams have also been reported in previous experiments [8, 9] and are attributed to the multiband superconductivity [9].

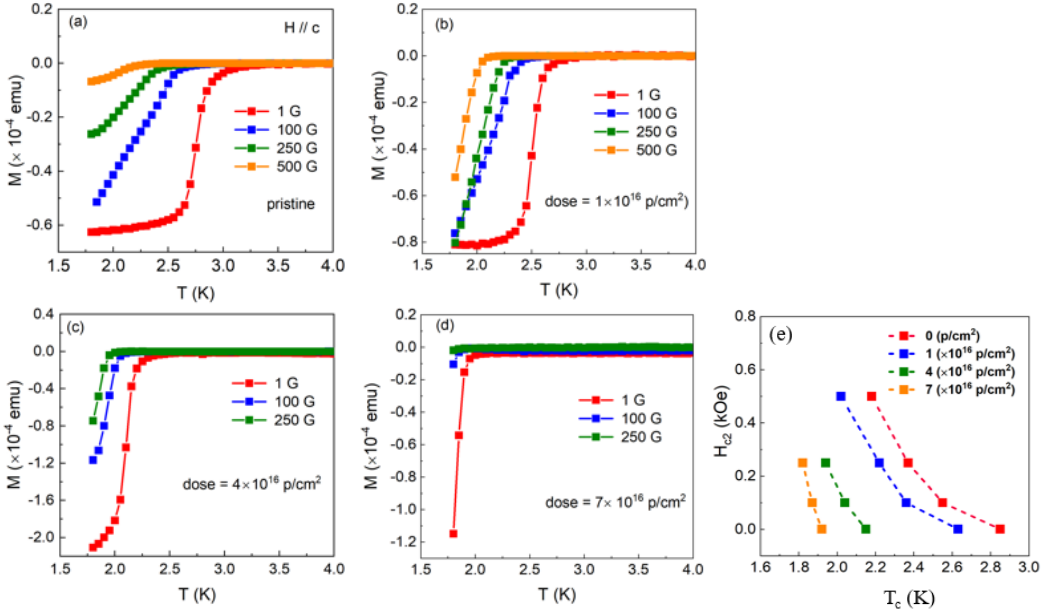

FIG. S3. (a) Temperature dependence of magnetization, $M(T)$ measured at various magnetic fields along the $c$-axis and as indicated with p-irradiation dose (a) pristine, (b) $1 \times 10^{16}$ p/cm$^2$, (c) $4 \times 10^{16}$ p/cm$^2$, (d) $7 \times 10^{16}$ p/cm$^2$. (e) Superconducting phase boundary for $H \parallel c$ constructed from the data in frames a-d. Following each dose, the upwards curvature is suppressed successively, consistent with the results obtained from the TDO measurement (see Fig. 3(d), main text).

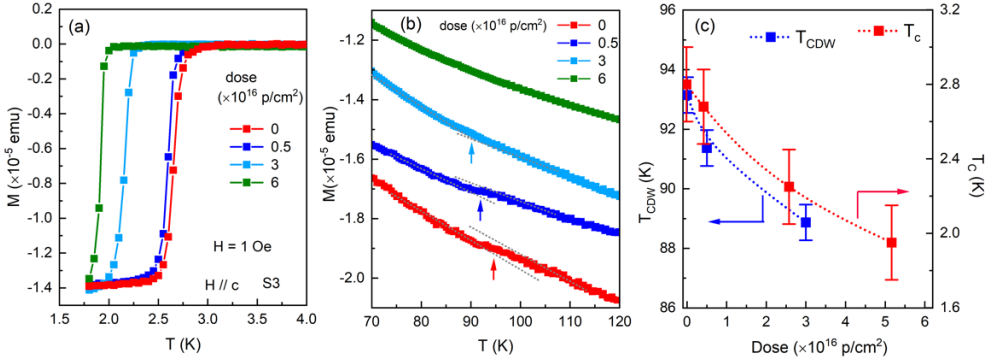

FIG. S4. Dependence of $T_c$ and $T_{CDW}$ of CsV$_3$Sb$_5$ sample S3 at indicated p-irradiation dose. (a)Temperature dependence of the magnetic moment, $M(T)$, at indicated doses measured by applying $H = 1$ Oe along $c$-axis displaying superconducting transition. (b)Temperature dependence $M(T)$ measured in a field of $H = 10$ kOe along $c$-axis displaying the CDW transition. (c) Transition temperatures $T_{CDW}$ and $T_c$ of sample S3 vs dose. These data are supplemented with the data on sample S2 in Fig. 3(c) with $T_{CDW}$ and $T_c$ both normalized to the values of the respective pristine sample. The error bars represent the width of the transitions.

**Temperature dependence of the TDO frequency of the pristine and irradiated CsV$_3$Sb$_5$**

The TDO technique provides highly sensitive measurements of the temperature and field dependence of the magnetic susceptibility of the sample, enabling the determination of the temperature dependence of the penetration depth and the superconducting phase diagram. Multiple CsV$_3$Sb$_5$ crystals were measured via the TDO technique. Figure S5 (a-f) shows the temperature dependence of the TDO frequency shift, $f_{TDO}(T)$, measured at various magnetic fields along the $c$-axis and $ab$ plane, and at indicated p-irradiation dose.

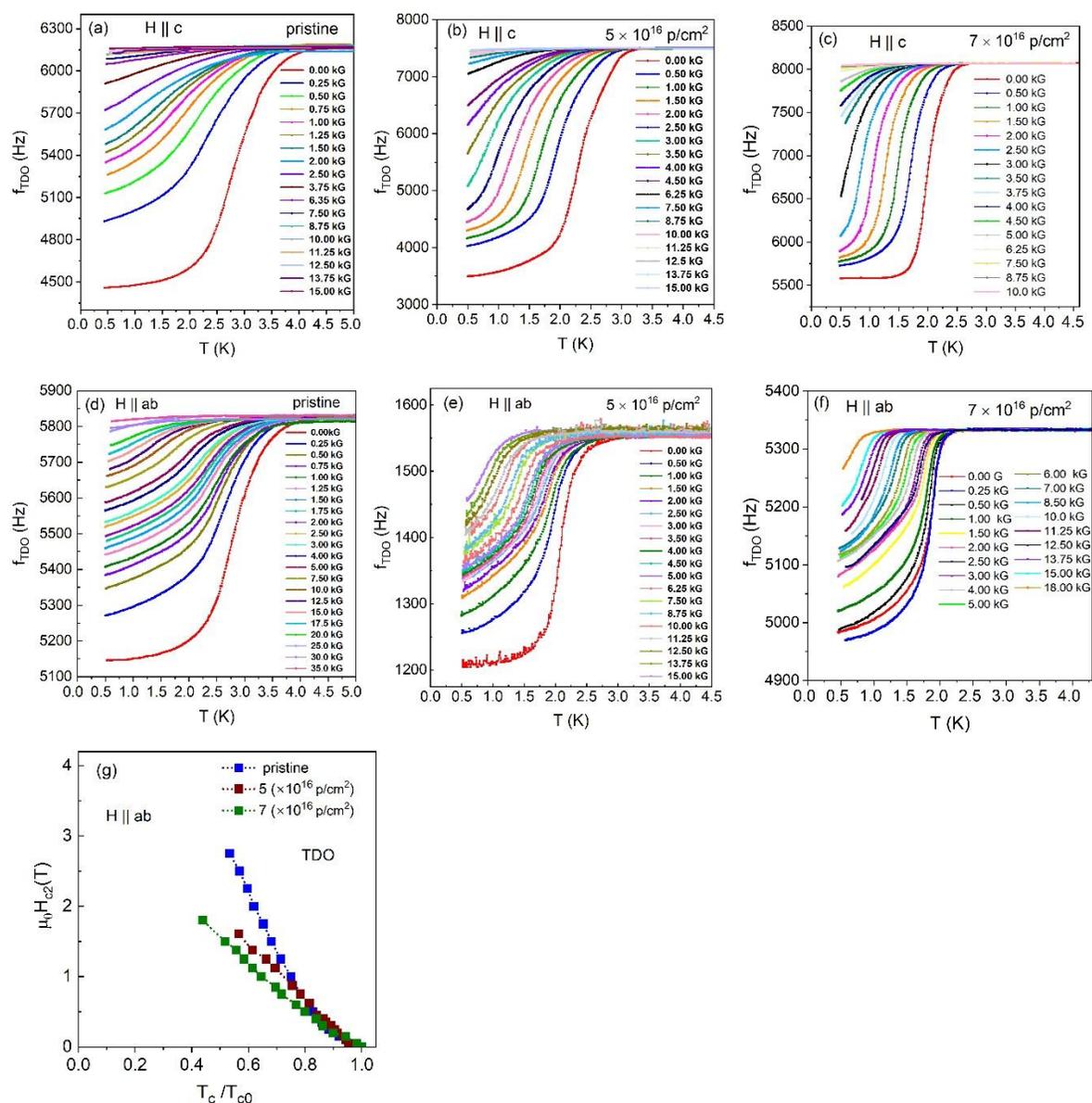

FIG. S5. (a-f) Temperature dependence of the TDO frequency measured at various magnetic fields applied along $c$-axis and ab plane and at indicated doses. These data are used to construct the phase diagram in Fig. 3(d) in the main text for $H \parallel c$ and Fig. S5 (g) for $H \parallel ab$.

To establish a consistent $T_c$ criterion between measurement techniques, we use the following procedure for the TDO data. First, we find the $\Delta f$ decrease from the temperature independent normal state behavior in zero applied field at the temperature equal to $T_{c,midpoint}$ from transport measurements. Here, we define this $\Delta f$ as our criterion for $T_c$ in all applied fields in TDO measurements. This approach forces $T_c$ from both TDO and transport in zero field to be identical, gives consistent agreement of the superconducting phase boundaries determined from each technique. Measuring $\Delta f(T)$ in various applied magnetic fields (see Fig. S5 (a-f)), the phase boundary $H_{c2}(T)$ vs $T$ is constructed for the pristine and irradiated sample for both field orientations. The upward curvature in the $H_{c2}(T)$ gradually suppresses with increasing proton dose (see Fig. 3(d), main text, for $H \parallel c$ and Fig. S5(g) for $H \parallel ab$), transitioning to the conventional behavior, consistent with the observations from transport and magnetization measurements.

**Estimation of dimensionless scattering parameter, $g^\lambda$**

To analyze the suppression of $T_c$ with irradiation dose, we estimate a dimensionless scattering parameter [7] $g^\lambda = \frac{\hbar \Delta \rho_0}{2\pi \mu_0 k_B T_{c,0} \lambda_0^2}$, yielding $g^\lambda \approx 1.05$ for $\Delta \rho_0 \approx 56.25$ ($\mu\Omega$ cm) at an irradiation dose of 6×10$^{16}$ p/cm$^2$ as obtained from the data in Fig. S1(c) and taking the zero-temperature penetration depth $\lambda_0 \approx 387\ nm$ [10]. As it proved challenging to maintain good electrical contacts after repeated irradiation, the systematic characterization of the dependence of $T_c$ and $T_{CDW}$ on irradiation dose was deduced from magnetization and susceptibility measurements on three crystals carried through a protocol of consecutive irradiations. While on sample S1 we have a direct measure of both $\Delta \rho_0$ and of $T_c/T_{c0}$, on samples S2 and S3 we obtained $T_c/T_{c0}$ at various irradiation doses from magnetic measurements. Therefore, to estimate $g^\lambda$, we assume a linear relation between $T_c/T_{c0}$ and $g^\lambda$ as indicated in Fig. S1(f). The results for samples S2 and S3 are summarized in Fig. 3 (c) (main text), showing $T_c$ suppression with increasing irradiation dose.

Potential scattering in isotropic superconductors should not suppress $T_c$ at all (Anderson theorem [11]). Thus, the observed $T_c$ suppression with increasing disorder indicates some degree of pair-breaking scattering and/or the presence of anisotropy. However, the rate of $T_c$ suppression with increasing disorder is much slower than that expected in the Abrikosov-Gor'kov (AG) model [12, 13] for dominant pair-breaking scattering indicating that the additional scattering due to irradiation-induced defects is predominantly not pair-breaking. In an anisotropic *s*-wave superconductor, impurity scattering can average out the anisotropic gap, making the gap effectively isotropic [14-16]. Subsequently, $T_c$ may initially drop rapidly with impurity density, but as the gap becomes more isotropic, the reduction of $T_c$ saturates. These expectations agree with observations shown in Fig. 3(c) (main text). We also notice that the variation of $T_c$ and $T_{CDW}$ of $CsV_3Sb_5$ induced by electron irradiation [14] is similar to our proton-irradiation results.

**In-plane anisotropy of the upper critical field**

We explored the in-plane anisotropy of $H_{c2}$ by mounting a crystal with its face parallel to the axis of the 9-T magnet at a sequence of angles phi between current and field directions. We check the alignment of the sample by monitoring its resistance in response to a small out-of-plane thereby effectively rotating the total applied field with respect to the sample plane as shown in the inset of Fig. S6 (a). In all cases, the misalignment was 1 deg or less. As seen in the inset of Fig. S6 (a), changes in the resistance arising from this misalignment are small enough as to rule out spurious effects in the angular dependence of $H_{c2}$.

The main panel of Fig. S6(a) shows the temperature dependence of the in-plane upper critical field for various angles between field and current. The overall shape of the phase boundary is similar to that shown in Fig. 1(d) of the Main Text. However, there is a noticeable angular dependence with period of 180 deg as discussed below. For fields below 1T we can perform complete high-resolution rotation diagrams. Alignment scans such as shown in Fig. S6(a) for two orthogonal directions define the orientation of the sample allowing to calculate the field components such that the resultant field always lies in the plane of the sample. The resulting data, Fig. S6(b), clearly reveal a 180-deg symmetry. This angular variation is oriented in such a way that the lowest resistance (highest $H_{c2}$ as defined by the 50%-criterion) corresponds to the magnetic field being aligned with the current. We therefore tentatively interpret this angular dependence as a Lorentz force effect. We note though that a two-fold in-plane symmetry in various properties of $CsV_3Sb_5$ has been reported, see for instance [17], which may be associated with the breaking of rotational symmetry in the CDW-phase and the formation of a nematic state [18]. A recent theoretical study employing the time dependent Ginzburg Landau approach [19] reveal pronounced anisotropy of the flux flow resistivity due to nematicity, in this case for a system of $C_4$ symmetry. This implies that the upper critical field as determined from resistive transitions may reflect nematic order. However, as our sample is macroscopic in scale and likely contains a pattern of CDW domains [20-24] we contend that nematic effects are largely averaged and that the dominant cause of 2-fold symmetry is the Lorentz force.

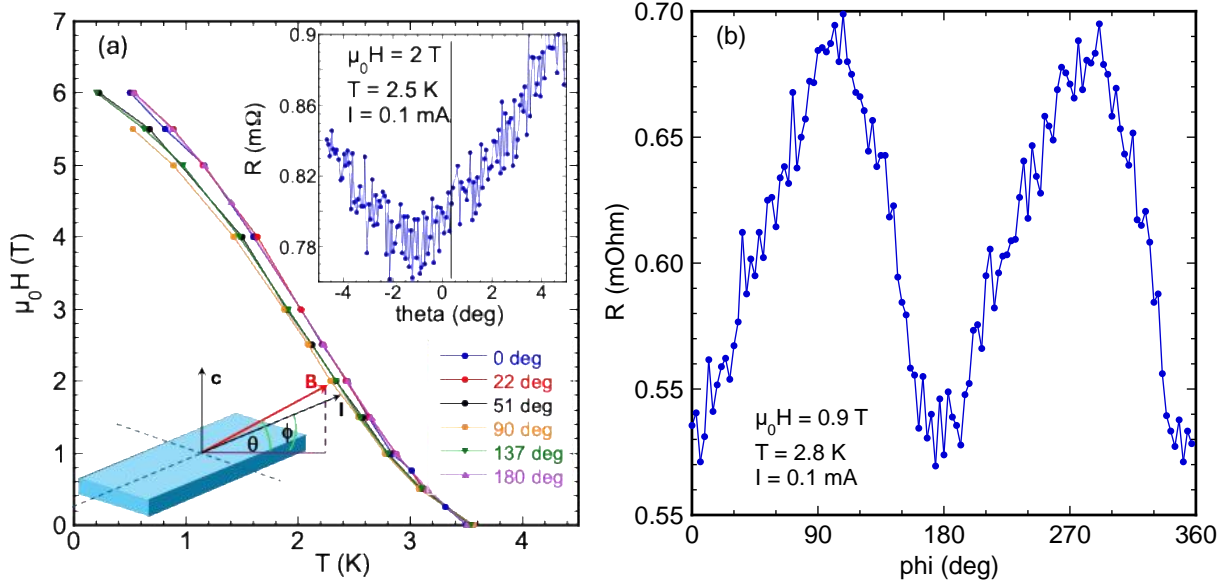

FIG. S6. (a) Temperature dependence of the in-plane $H_{c2}$ for various angles between field and current directions. The top inset shows an alignment scan where angle theta measures the angle of the applied field with respect to the sample plane. The geometry is shown in the bottom inset. (b) In-plane angular dependence of the sample resistance in a field of 0.9 T at 2.8 K. At phi = 0, current and applied field are parallel.

**Upper critical fields for two-gap layered superconductor**

To model the $H_{c2}(T)$ data of CsV$_3$Sb$_5$ (see Fig. 1(c)), we consider properties of multiple-band BCS layered superconductors including gap equation and upper critical field without paramagnetic effect [25]. The general gap equation for a multiple-band BCS model is

$$\Delta_\alpha = \Lambda_{\alpha\beta} \int_0^{\omega_D} d\varepsilon \frac{\Delta_\beta}{\sqrt{\varepsilon^2+\Delta_\beta^2}} \tanh \frac{\sqrt{\varepsilon^2+\Delta_\beta^2}}{2T} \tag{A1}$$

with the coupling-constant matrix $\Lambda_{\alpha\beta} = U_{\alpha\beta} N_\beta$, where $U_{\alpha\beta}$ are pairing interaction and $N_\beta$ are partial densities of states (DOSs). In particular, for open Fermi surface and parabolic spectrum, the DOS is given by $N_\alpha = \frac{m_\alpha}{2\pi\hbar^3 s}$ where $s = 2\pi T_c \rho$ [26]. Superconducting instability develops at the eigenvector $\Omega_\alpha$ of the matrix $\Lambda_{\alpha\beta}$ which corresponds to the largest eigenvalue $\lambda$ giving the effective coupling constant, $\sum_\beta \Lambda_{\alpha\beta} \Omega_\beta = \lambda \Omega_\beta$. This eigenvector is normalized by the condition $\sum_\alpha n_\alpha \Omega_\alpha^2 = 1$. To evaluate the temperature dependence of c-axis upper-critical field, we consider the multiple-band Eilenberger equations with finite Zeeman term and exchange field as

$$\Delta_\alpha(\mathbf{r}) = 2\pi T \sum_{\beta,\omega} \Lambda_{\alpha\beta} \langle f_\beta(\mathbf{r},\mathbf{k},\omega) \rangle_\beta \tag{A2}$$

$$(2\omega + \mathbf{v}_\alpha \cdot \mathbf{\Pi}) f_\alpha = 2\Delta_\alpha, \tag{A3}$$

where $f_\alpha(\mathbf{r},\mathbf{k},\omega)$ are the anomalous Green's functions, $\mathbf{\Pi} = \nabla + 2\pi i \mathbf{A}/\Phi_0$, $\omega = 2\pi T \left(n + \frac{1}{2}\right)$ is the Matsubara frequency, and $\langle \ldots \rangle_\beta$ implies averaging over the $\beta$ Fermi sheet. A single-channel approximation corresponding to the presentation $\Delta_\alpha(\mathbf{r}) = \Psi(\mathbf{r})\Omega_\alpha$ yields

$$\Psi(\mathbf{r}) = 2\pi T \sum_{\alpha,\beta,\omega} n_\alpha \Omega_\alpha \Lambda_{\alpha\beta} \langle f_\beta(\mathbf{r},\mathbf{k},\omega) \rangle_\beta \tag{A4}$$

$$(2\omega + \mathbf{v}_\alpha \cdot \mathbf{\Pi}) f_\alpha = 2\Psi(\mathbf{r})\Omega_\alpha \tag{A5}$$

The first equation can be regularized using equation for $T_c$, $\Omega_\alpha = 2\pi T_c \sum_{\beta,\omega} \Lambda_{\alpha\beta} \frac{\Omega_\beta}{\omega}$, as

$$\ln \frac{T_c}{T} \Psi(\mathbf{r}) = -2\pi T \sum_{\beta,\omega} n_\beta \Omega_\beta \left\langle f_\beta(\mathbf{r},\mathbf{k},\omega) - \frac{\Psi(\mathbf{r})}{\omega}\Omega_\beta \right\rangle_\beta. \tag{A6}$$

Equation (A5) can be formally solved as

$$f_\alpha = 2\int_0^\infty d\rho \exp[-\rho(2\omega + v_\alpha \cdot \mathbf{\Pi})]\Psi(\mathbf{r})\Omega_\alpha.$$

Substituting this solution the self-consistency equation (A6) and performing summation over $\omega$,

$\sum_\omega \exp(-2\rho\omega) = \sum_{n=0}^\infty \exp(-4\pi\rho T(n+1/2)) = \frac{1}{2\sinh(2\pi\rho T)}$, we rewrite Eq. (A6) as

$$\ln \frac{T_c}{T} \Psi(\mathbf{r}) = \sum_\alpha w_\alpha \int_0^\infty \frac{2\pi T d\rho}{\sinh(2\pi\rho T)} (1 - \langle \exp(-\rho v_\alpha \cdot \mathbf{\Pi})\rangle_\alpha)\Psi(\mathbf{r}), \tag{A7}$$

$$\text{where } w_\alpha = n_\alpha \Omega_\alpha^2 = \frac{n_\alpha \Delta_\alpha^2}{\sum_\beta n_\beta \Delta_\beta^2}$$

(A8)

are the band widths. In particular for two bands, we can write the band weights as

$w_1 = \frac{1}{1+r_n r_\Delta}$ and $w_2 = \frac{r_n r_\Delta}{1+r_n r_\Delta}$, where $r_\Delta = \Delta_2^2/\Delta_1^2$ and $r_n = N_2/N_1$ (ratio of density of states). For quasi-2D model, $r_n = r_m = m_2/m_1$ (ratio of effective masses). Equation (A7) has analytical solution only in few special situations such as spherical Fermi surfaces. Also, it always can be solved analytically near $T_c$, where it can be reduced to the Schrodinger equation for a particle in magnetic field and $\Psi(\mathbf{r})$ is given by the lowest-Landau-level wave function. In other cases, one can either solve it numerically or rely on approximations. In many cases, an approximate simpler calculation is preferable because of uncertainties of the model parameters. The simplest approximation is to project Eq. (A7) to the lowest-Landau-level wave function in the whole temperature range. For field along $z$ axis, this approximation gives the following equation for the upper critical field

$$\ln\frac{T_c}{T} = \sum_\alpha w_\alpha \int_0^\infty \frac{2\pi T d\rho}{\sinh(2\pi\rho T)}\left[1 - \left\langle\exp\left(-\frac{\pi}{2}v_\alpha^2\frac{H\rho^2}{\Phi_0}\right)\right\rangle_\alpha\right]. \tag{A9}$$

For numerical calculations, it is convenient to introduce the reduced variables, $t = T/T_c$, $s = 2\pi T_c \rho$,

$$\mathcal{H} = \frac{H}{H_{z0}}, H_{z0} = \frac{16\pi\Phi_0 T_c^2}{\hbar^2 v_{GL}^2}, \quad v_{GL}^2 = \sum_\alpha w_\alpha \langle v_\alpha^2\rangle_\alpha, \tag{A10}$$

With the reduced parameters, Eq. (A9) simplifies to

$$\ln\frac{1}{t} = \sum_\alpha w_\alpha \int_0^\infty \frac{tds}{\sinh(ts)}\left(1 - \left\langle\exp\left(-2\mathcal{H}\frac{v_\alpha^2}{v_{GL}^2}s^2\right)\right\rangle_\alpha\right) \tag{A11}$$

One can also derive an equivalent form without 'zero over zero' behavior in the integral using $\frac{tds}{\sinh(ts)} = d\ln\tanh\left(\frac{ts}{2}\right)$ and integrating by parts

$$-\ln t = \int_0^\infty ds \ln\tanh\left(\frac{ts}{2}\right)\sum_\alpha w_\alpha \frac{d}{ds}\left\langle\exp\left(-2\mathcal{H}\frac{v_\alpha^2}{v_{GL}^2}s^2\right)\right\rangle_\alpha. \tag{A12}$$

We now apply the above general results to the case of finite interlayer tunneling. We assume the simplest dispersion relations

$$\varepsilon(p) = \frac{p_x^2 + p_y^2}{2m} + 2t_z\cos(p_z d) - \varepsilon_F \quad (d \text{ is the c-axis lattice paramerter}) \tag{A13}$$

with $2t_z < \varepsilon_F$ giving $\quad p_F(p_z) = \sqrt{2m(\varepsilon_F - 2t_z\cos(p_z d))}, v_F(p_z) = v_F\sqrt{1 - \frac{2t_z}{\varepsilon_F}\cos(p_z d)},$

$$\tag{A14}$$

$$v_x(p_z,\phi) = v_F(p_z)\cos\phi, v_z = -2t_z d\sin(p_z d) \tag{A15}$$

where $v_F = \sqrt{2\varepsilon_F/m}$ and $\bar{v}_x^2 = \frac{v_F^2}{2}, \bar{v}_z^2 = 2t_z^2 d^2$ \hfill (A16)

Furthermore, the average $\left\langle\exp\left(-2\mathcal{H}\frac{v^2}{v_{GL}^2}s^2\right)\right\rangle = \int_0^\pi \frac{dq}{\pi}\exp\left[-2\mathcal{H}\frac{v_F^2}{v_{GL}^2}s^2\left(1 - \frac{2t_z}{\varepsilon_F}\cos q\right)\right]$

$= \exp\left(-2\mathcal{H}\frac{v_F^2}{v_{GL}^2}s^2\right)I_0\left(2\mathcal{H}\frac{v_F^2}{v_{GL}^2}\frac{2t_z}{\varepsilon_F}s^2\right)$, where $\int_0^\pi \frac{dx}{\pi}\exp(a\cos x) = I_0(a)$. Then Eq. (A12) becomes

$$-\ln t = \int_0^\infty ds \ln\tanh\left(\frac{ts}{2}\right)\sum_\alpha w_\alpha \frac{d}{ds}\exp\left(-2\mathcal{H}\frac{v_{F,\alpha}^2}{v_{GL}^2}s^2\right)I_0\left(2\mathcal{H}\frac{v_{F,\alpha}^2}{v_{GL}^2}\frac{2t_{z,\alpha}}{\varepsilon_{F,\alpha}}s^2\right)$$
$$\tag{A17}$$

For a two-band superconductor, introducing the velocity ratio $r_v = v_{F,2}^2/v_{F,1}^2$, we can present

$$\frac{v_{F,1}^2}{v_{GL}^2} = \frac{1}{w_1+w_2 r_v}, \quad \text{and} \quad \frac{v_{F,2}^2}{v_{GL}^2} = \frac{r_v}{w_1+w_2 r_v},$$

where $w_2 = 1 - w_1$. For $t_{z,\alpha} \ll \varepsilon_{F,\alpha}$, the shape of the $H_{c2}(T)$ curve is determined by two parameters $w_1$ and $r_v$. Introducing the reduced function, for out-of-plane $H_{c2}$, one can write

$$\mathcal{G}_z(t,\tau_z,\zeta) = -\ln\frac{1}{t} + \int_0^\infty ds \ln\tanh\left(\frac{ts}{2}\right)\frac{d}{ds}[\exp(-2\zeta s^2)I_0(4\zeta\tau_z s^2)], \quad (A18)$$

with $\zeta = \mathcal{H}\frac{v_a^2}{v_{GL}^2}$ and $\tau_z = t_z/\varepsilon_F$. In terms of these functions, Eq. (A17) can be written as

$$\sum_\alpha w_\alpha \mathcal{G}_z\left(t, \frac{t_{z,\alpha}}{\varepsilon_{F,\alpha}}, \mathcal{H}\frac{v_a^2}{v_{GL}^2}\right) = 0. \quad (A19)$$

For field along *ab*-plane, following similar procedure for in- plane $H_{c2}$ one can derive

$$\mathcal{G}_y(t,\tau_z,\zeta_x,\zeta_z) = -\ln\frac{1}{t} + \int_0^\infty \frac{t ds}{\sinh(ts)}\int_0^\pi \frac{dq}{\pi}$$
$$\{1 - \exp(-2\zeta_z\sin^2 q s^2)\exp[-\zeta_x(1-2\tau_z\cos q)s^2]I_0[\zeta_x(1-2\tau_z\cos q)s^2]\}$$

(A20)

with $\zeta_x = \mathcal{H}\frac{v_{F,\alpha}^2}{v_{GL}^2}$, $\zeta_z = \mathcal{H}\frac{t_{z,\alpha}^2}{\bar{t}_z^2}$, $\bar{t}_z^2 = \sum_\alpha w_\alpha t_{z,\alpha}^2$, we rewrite this equation as

$$\sum_\alpha w_\alpha \mathcal{G}_y\left(t, \frac{t_{z,\alpha}}{\varepsilon_{F,\alpha}}, \mathcal{H}\frac{v_{F,\alpha}^2}{v_{GL}^2}, \mathcal{H}\frac{t_{z,\alpha}^2}{\bar{t}_z^2}\right) = 0$$

(A21)

We then fit the experimental data in Fig. 1(c) (main text) using Eqs. (A18) and (A19) for $H_{c2,c}(T)$ and using Eqs. (A19) and (A20) for $H_{c2,ab}(T)$ which yields $w_1 = 0.46$, $r_v = 57.6$, $t_{z1}/\varepsilon_{F1} = 0.057$, $t_{z2}/\varepsilon_{F2} = 0.133$ and $H_{c2,c}(0) = 1.1$ T; $H_{c2,ab}(0) = 6.0$ T. The large value of the Fermi-velocity ratio squared $r_v$ is required to reproduce the pronounced experimental upward curvature of $H_{c2,c}(T)$. The resulting fit is presented in the Fig. 1(c) of the main text.

**Upper critical field for open Fermi surface close to van Hove singularities**

We evaluate shapes of the temperature dependences of the upper critical field for Fermi surfaces close to van Hove singularities within quasiclassical approximation. The key feature of the electronic spectrum of the layered kagome superconductor CsV$_3$Sb$_5$ is the close proximity of several Fermi-surface sheets to van Hove singularities. Such pockets are characterized by strong variations of Fermi velocities which significantly affects the shapes of the temperature dependences of the upper critical fields. In the two-dimensional case, the influence of van Hove singularities on the upper critical field was investigated in Refs. [27-29]. To qualitatively model the shapes of the upper critical field in CsV₃Sb₅, we use a minimum single-band model (described in Ref. [30]) and and evaluate the upper critical fields for the case when the Fermi level crosses the van Hove energy. A sketch of a two-dimensinal cross section of the hexagonal Fermi surface is illustrated in Fig. S7(a).

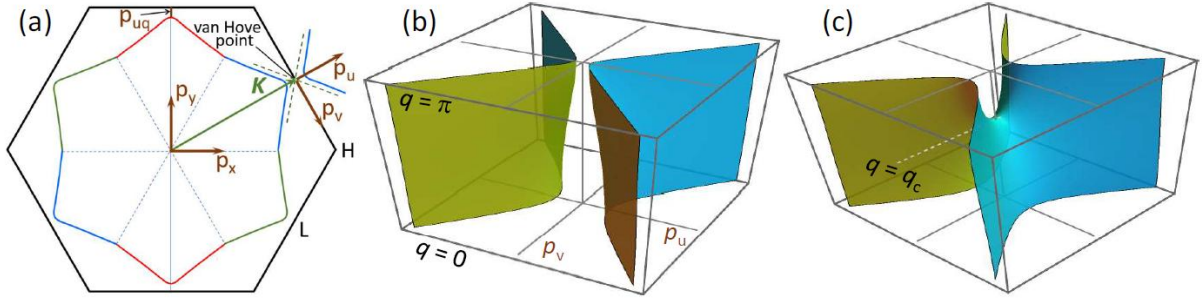

FIG. S7. (a) Finite $q$-cross section of the hexagonal Fermi pocket with corners close to van Hove singularities [30]. (b) Three-dimensional view of the Fermi-surface fragment near the van Hove line in the case when the Fermi level does not cross the van Hove energy. (c) Fermi-surface section in the case when the Fermi level crosses the van Hove energy at $q = q_c$.

**Model for electronic spectrum**

To qualitatively model the behavior of $H_{c2}(T)$ in CsV₃Sb₅, we use a minimum single-band model with the electronic spectrum in the form: $\varepsilon(\boldsymbol{p}, q) = \varepsilon_{2D}(\boldsymbol{p}) + \varepsilon_z(q)$ where $\boldsymbol{p}$ and $q$ are in-plane and $c$-axis momenta, $\varepsilon_{2D}(p)$ is the spectrum for a single layer and $\varepsilon_z(q) = -2t_z \cos q$ is the tunneling term. The same model has been used in our previous work [30] to interpret anomalous magnetotransport in CsV₃Sb₅. We assume that the Fermi level is close to the van Hove singularities located near six wave vectors $\boldsymbol{K}_j$ with

$$K_{j,x} = K \cos\left(-\frac{\pi}{6} + \frac{\pi}{3}j\right), \quad K_{j,y} = K \sin\left(-\frac{\pi}{6} + \frac{\pi}{3}j\right), \tag{B1}$$

where $j = 1, 2, \ldots, 6$, $K = 2\pi/(\sqrt{3}a)$ is a half of the basic reciprocal-lattice vector for the unfolded Brillouin zone, and $a$ is the lattice parameter (5.45 Å for CVS). In the vicinity of these wave vectors, the spectrum has a saddle-point of the form

$$\varepsilon_{2D,j}(\boldsymbol{p}) = \frac{p_u^2}{2m_u} - \frac{p_v^2}{2m_v}, \tag{B2}$$

where $p_u$ and $p_v$ are the components of the vector $\boldsymbol{p} - \boldsymbol{K}_j$ along and perpendicular to $\boldsymbol{K}_j$, respectively, see Fig. S8(a), $m_u$ and $m_v$ are effective masses, and $m_v/m_u = r_m$ is their ratio. The shape of the Fermi surface is determined by equation $\varepsilon(\boldsymbol{p}, q) = \varepsilon_{vH}$, where $\varepsilon_{vH}$ is the distance between the Fermi level and van Hove energy for $t_z = 0$. Depending on the relation between $t_z$ and $\varepsilon_{vH}$, the Fermi level may either cross or not cross the van Hove energy as function of $q$. The first and second scenario is realized for $2|t_z| > \varepsilon_{vH}$ and $2|t_z| < \varepsilon_{vH}$, respectively and the corresponding Fermi surfaces are illustrated in Fig. S7(b) and (c).

***c*-axis upper critical field for the case of Fermi level crossing van Hove energy**

Our analysis of the *c*-axis upper critical field is based on the quasiclassical result in Eq. (2) (see main text). In our approximate calculations, we assume the hyperbolic electronic spectrum in Eq. (B2) and will cut off the integrations along the Fermi surface at $|p_u| = K_c$, where the cutoff wave vector $K_c$ is close to the half of the reciprocal-lattice vector $K$. For definiteness, we consider the case of the Fermi level crossing the van Hove energy. Since the topology of the Fermi surface changes the Fermi level crossing the van Hove energy at $q = q_c$, the averaging with respect the to *c*-axis wave vector has to be split into two regions

$$\langle G(v_x, v_y)\rangle = \frac{1}{\int_0^\pi \frac{dq}{\pi} \oint \frac{dp}{v}} \left[ \int_0^{q_c} \frac{dq}{\pi} \oint \frac{dp}{v} G(v_x, v_y) + \int_{q_c}^\pi \frac{dq}{\pi} \oint \frac{dp}{v} G(v_x, v_y) \right]. \tag{B3}$$

The Fermi surface averaging for the spectrum in Eq. (B2) can be conveniently carried out using hyperbolic parametrization [25]. For example, for $q < q_c$ the momentum components are parameterized as $p_u = p_{u0}\sqrt{1 - \vartheta_z(q)} \cosh\eta$ and $p_v = p_{v0}\sqrt{1 - \vartheta_z(q)} \sinh\eta$ with $p_{u0} = \sqrt{2m_u \varepsilon_{vH}}$ and $p_{v0} = \sqrt{2m_v \varepsilon_{vH}}$. The mean-square average of the *x*-component of velocity can be evaluated using the hyperbolic parametrizations as

$$\bar{v}_x^2 = \frac{v_{v0}^2 + v_{u0}^2}{4\bar{\eta}_b} \int_0^\pi \frac{dq}{\pi} \left( \frac{\sinh(2\eta_b)}{2} + \delta_q \eta_b \alpha_v \right) |1 - \vartheta_z(q)| \tag{B4}$$

where the $q$-dependent parameter $\eta_b \approx \ln(2\kappa_c/\sqrt{|1 - \vartheta_z(q)|})$, $\kappa_c = K_c/p_{u0}$, $p_{u0} = \sqrt{2m_u \varepsilon_{vH}}$, $\vartheta_z(q) = \varepsilon_z(q)/\varepsilon_{vH}$, $\bar{\eta}_b = \int_0^\pi \frac{dq}{\pi} \eta_b(q) = \ln(2w_z \kappa_c)$, $w_z = \sqrt{\varepsilon_{vH}/|t_z|}$. $v_{u0} = \sqrt{2\varepsilon_{vH}/m_u}$, $v_{v0} = \sqrt{2\varepsilon_{vH}/m_v}$, $\alpha_v = (r_m - 1)/(r_m + 1)$, and $\delta_q = \text{sign}(q_c - q)$. In the case $\eta_b \gg 1$, we can approximately obtain

$$\bar{v}_x^2 = \frac{v_{v0}^2 + v_{u0}^2}{2r_x} \tag{B5}$$

with

$$r_x \approx \frac{2}{\kappa_c^2} \ln(2w_z \kappa_c) \tag{B6}$$

Here we use the analytical result for the integral

$$\int_0^\pi \frac{dq}{\pi} \ln|1 + a\cos q| = \ln\left(\frac{|a|}{2}\right), \text{ for } a > 1. \tag{B7}$$

Using the hyperbolic parametrizations, we can approximately perform the Fermi-surface averaging in Eq. (2) for fixed $q$,

$$\left\langle \exp\left[-\left(\frac{v_x^2}{\bar{v}_x^2} + \frac{v_y^2}{\bar{v}_y^2}\right) hs^2\right]\right\rangle$$

$$= \frac{1}{\bar{\eta}_b} \langle \int_0^{\eta_b} d\eta \, \exp\{-r_x[\delta_q \alpha_v + \cosh(2\eta)]|1 - \vartheta_z(q)|hs^2\}\rangle_q.$$

This allows us to rewrite the equation for the upper critical field (Eq. (2), main text) as

$$-\ln t = \int_0^\infty \frac{tds}{\sinh(ts)} \left\{ 1 - \frac{1}{\bar{\eta}_b} \langle \int_0^{\eta_b} d\eta \, \exp\{-r_x[\delta_q \alpha_v + \cosh(2\eta)]|1 - \vartheta_z(q)|hs^2\}\rangle_q \right\} \tag{B8}$$

or

$$-\ln t = \frac{1}{\bar{\eta}_b} \int_0^\infty ds \, \ln \tanh\left(\frac{ts}{2}\right) \frac{d}{ds} \langle \int_0^{\eta_b} d\eta \, \exp\{-r_x[\delta_q \alpha_v + \cosh(2\eta)]|1 - \vartheta_z(q)|hs^2\}\rangle_q. \tag{B9}$$

The case $2t_z < \varepsilon_{vH}$ can be treated using the same equations by setting $q_c = \pi$.

Near $T_c$, we can expand the right hand side of Eq. (2) (main text) with respect to $h$, which yields the Ginzburg-Landau linear bahavior $h \simeq h_{GL}(1-t)$ and, using the integral $\int_0^\infty \frac{s^2 ds}{\sinh s} = \frac{7\varsigma(3)}{2}$, we obtain, $h_{GL} = \frac{1}{7\varsigma(3)}$. In real units, this result approximately corresponds to

$$H_{c2,z}(t) \simeq \frac{32\pi}{7\varsigma(3)} \frac{T_c^2 \Phi_0}{\hbar^2 \kappa_c^2 (v_{u0}^2 + v_{v0}^2)} \ln\left(2\kappa_c \sqrt{\frac{\varepsilon_{vH}}{|t_z|}}\right)(1-t). \tag{B10}$$

Note that, contrary to the 2D case, the upper critical field linear slope $dH_{c2}/dT$ approaches finite value for $\varepsilon_{vH} \to 0$. In the low-temperature limit, using the integral $\int_0^\infty ds \, \ln\left(\frac{s}{2}\right) \frac{d}{ds} \exp(-as^2) = \frac{1}{2}(\gamma_E + \ln 4a)$, where $\gamma_E = 0.5772$ is the Euler constant, Eq. (B9) can be written as

$$\gamma_E + \ln(4r_x h) + \frac{1}{\bar{\eta}_b} \langle \int_0^{\eta_b} d\eta \, \ln[(\delta_q \alpha_v + \cosh(2\eta))]\rangle_q + \langle \ln|1 - \vartheta_z(q)|\rangle_q = 0. \tag{B11}$$

In the limit $\eta_b \gg 1$ this equation can be approximately transformed to the following form

$$\gamma_E + \ln(4r_x h) + \ln \kappa_c + \frac{1}{2}\ln\left(\frac{|t_z|}{\varepsilon_{vH}}\right) = 0, \tag{B12}$$

which gives the zero-temperature upper critical field in the reduced form

$$h_{c2,z}(0) \simeq \frac{\exp(-\gamma_E)}{4}\sqrt{\frac{\varepsilon_{vH}}{|t_z|}}\frac{1}{r_x\kappa_c}. \tag{B13}$$

In real units, this result corresponds to

$$H_{c2,z}(0) \simeq 4\pi\exp(-\gamma_E)\sqrt{\frac{\varepsilon_{vH}}{|t_z|}}\frac{T_c^2\Phi_0}{\hbar^2\kappa_c(v_{v0}^2+v_{u0}^2)}, \tag{B14}$$

The ratio of the zero-temperature and GL extrapolated values

$$\frac{H_{c2,z}(0)}{H_{GL,z}} \simeq \frac{7\varsigma(3)}{16}\exp(-\gamma_E)\sqrt{\frac{\varepsilon_{vH}}{|t_z|}}\frac{2\kappa_c}{\ln\left(2\kappa_c\sqrt{\frac{\varepsilon_{vH}}{|t_z|}}\right)}$$

$$= \frac{7\varsigma(3)}{8}\exp(-\gamma_E)\sqrt{\frac{K_c^2}{2m_u|t_z|}}\frac{1}{\ln\left(2\sqrt{\frac{K_c^2}{2m_u|t_z|}}\right)} \tag{B15}$$

We can see that in the case of Fermi level crossing van Hove point the temperature dependence of $c$-axis upper critical field also has strong upward curvature due to the inequality $t_z \ll K_c^2/2m_u$.

**In-plane upper critical field**

For the magnetic field along the $y$ axis, the derivation based on general quasiclassical formula in Eq. (2) (main text) gives the following presentation for the reduced in-plane upper critical field $h$

$$-\ln t = \int_0^\infty \frac{tds}{\sinh(ts)}\left[1 - \frac{1}{\bar{\eta}_b}\int_0^\pi \frac{dq}{\pi}\int_0^{\eta_b} d\eta \left\langle \exp\left[-\left(g_{x,j}(\eta) + 2\sin^2 q\right)hs^2\right]\right\rangle_j\right], \tag{B16}$$

where $g_{x,j}(\eta)$ is the ratio $v_x^2/\bar{v}_x^2$ in the vicinty of the van Hove vector $\boldsymbol{K}_j$,

$$g_{x,j}(\eta) = \begin{pmatrix} 2r_x\frac{(\sinh\eta c_j - \sqrt{r_m}\cosh\eta s_j)^2}{1+r_m}(1-\vartheta_z(q)), & \text{for}\vartheta_z(q) < 1 \\ 2r_x\frac{(\cosh\eta c_j + \sqrt{r_m}\sinh\eta s_j)^2}{1+r_m}(\vartheta_z(q) - 1), & \text{for}\vartheta_z(q) > 1 \end{pmatrix}, \tag{B17}$$

$c_j = \cos(\pi j/3)$, and $s_j \equiv \sin(\pi j/3)$, and $\langle ... \rangle_j$ means averaging over the six van Hove wave vectors $\boldsymbol{K}_j$. Near the transition temperature, the result for the linear dependence is

$$H_{c2,y}(t) = H_{GL,y}(1-t), \tag{B18}$$

$$H_{GL,y} = \frac{16\pi}{7\sqrt{2}\zeta(3)}\frac{T_c^2\Phi_0}{\hbar\sqrt{v_{v0}^2+v_{u0}^2}t_z d}\frac{\sqrt{\ln(2w_z\kappa_c)}}{2\kappa_c}. \tag{B19}$$

This slope is also determined by the behavior of the in-plane velocities far away from the van Hove points. In the limit $\varepsilon_{vH} \to 0$ the slope diverges even slower than for the c-axis case, as the square root of the logarithm. The GL anisotropy factor for our model we can evaluate as

$$\gamma_{GL} = \frac{\sqrt{v_{v0}^2 + v_{u0}^2}\kappa_c}{2\sqrt{2\ln(2w_z\kappa_c)}t_z d/\hbar}. \tag{B20}$$

Its numerical value depends on details of the in-plane velocity behavior far away from the van Hove singularities which determines the cut off wave vector $K_c$. In the zero-temperature limit, the reduced in-plane upper-critical field can be evaluated as

$$H_{c2,y}(0) = 2\sqrt{2}\pi \exp(-\gamma_E) \frac{T_c^2 \Phi_0}{\hbar\sqrt{v_{v0}^2+v_{u0}^2}t_z d} \frac{\sqrt{\ln(2w_z\kappa_c)}}{2\kappa_c} \exp(-L_{av}) \tag{B21}$$

with $L_{av} = \left\langle \ln\left(\frac{v_x^2}{\bar{v}_x^2} + \frac{v_z^2}{\bar{v}_z^2}\right)\right\rangle = \frac{1}{\bar{\eta}_b}\int_0^\pi \frac{dq}{\pi}\frac{1}{6}\sum_j \int_0^{\eta_b} d\eta \ln(g_{x,j}(\eta) + 2\sin^2 q)$. In the limit $\eta_b \gg 1$, we derived the following estimate $L_{av} \approx -\ln 2 + A_v 2\sqrt{2}/\bar{\eta}_b$ with $A_v = \frac{1}{6}\sum_{j=1}^{6}\left|\cos\left(\frac{\pi}{3}j + \beta_v\right)\right|$ and $\beta_v = \arcsin(\sqrt{r_m}/\sqrt{1+r_m})$. The ratio of the zero-temperature and GL values for the in-plane direction can be evaluated as

$$\frac{H_{c2,y}(0)}{H_{GL,y}} = \frac{7\zeta(3)}{4}\exp(-\gamma_E - L_{av}). \tag{B22}$$

This ratio is typically larger than one implying upward curvature. However, contrary to the ratio for the c-axis direction (Eq. B15), it does not contain a large parameter, meaning that for the in-plane direction upward curvature is weaker (see. FIG. 1(d) in the main text).

**Representative temperature dependences of the upper critical field**

Full temperature dependence of the upper critical field can be obtained by numerical solution of Eqs. (B8) and (B16). Figure S8(lower panel) shows plots of the reduced z-axis upper critical field for $k_c = 10$ and two values of $t_z/\varepsilon_{vH}$ representing the cases of Fermi level separated from the van Hove energy ($t_z/\varepsilon_{vH} = 0.45$) and crossing the van Hove energy ($t_z/\varepsilon_{vH} = 0.6$). The latter shape approximately reproduces the observed behavior of the c-axis upper critical field in $CsV_3Sb_5$. We can conclude that the upward curvature of $H_{c2}(T)$ dependence for the in-plane direction is noticeably weaker than for the *c*-axis direction. This feature may be considered as a hallmark of strong influence of van Hove singularities on the upper critical field. Fig. S8(top) shows plots of reduced $y$-axis upper critical field. The plots are made for $k_c = 10$ and two values of $t_z/\varepsilon_{vH}$ representing the case of Fermi level separated from the van Hove energy ($t_z/\varepsilon_{vH} = 0.45$) and crossing van Hove energy ($t_z/\varepsilon_{vH} = 0.6$).

The plots show upward curvature but it is somewhat weaker than the experimental results (see Fig. 1(c-d), main text). We notice that the upwards curvature is suppressed due to gap anisotropy. To regain a good fit to the data, one needs to increase the value of κ implying that the Fermi surface comes closer to the vHs and that anisotropy of the Fermi velocity around the Fermi surface increases.

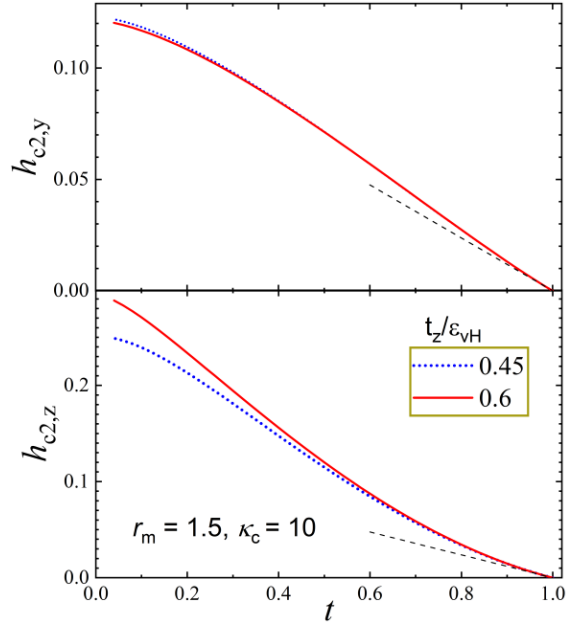

FIG. S8. Plots of reduced upper critical fields for representative parameters.

**Role of gap anisotropy**

We also investigated the influence of gap anisotropy on the $H_{c2}(T)$ shapes. We consider the case when the gap has maximum at the van Hove points and, for a qualitative assessment, we assume a simple dependence of the gap on the momentum components of the form

$$\Delta(p_v, p_u) = \Delta_0\sqrt{1 - c_u p_u^2/K_c^2 - c_v p_v^2/K_c^2}, \quad (B23)$$

which is determined by the two phenomenological constants $c_u$ and $c_v$. This form allows us to treat both cases of the Fermi level intersecting and not intersecting the van Hove point as a function of q. In our model, this shape of gap anisotropy corresponds to the following ratio of minimum and maximum gap, $\Delta_{min}/\Delta_{max} \approx \sqrt{1 - c_u - r_m c_v}$.

For a finite gap anisotropy, we modify Eqs. (B8) and (B16) by introducing an additional factor

$$\Omega^2 \propto [\Delta(p_v, p_u)]^2 \propto 1 - (c_u p_u^2 + c_v p_v^2)/K_c^2,$$

normalized by the condition $\langle\Omega^2\rangle = 1$. Such gap anisotropy increases the relative contribution of the Fermi-surface regions close to the van Hove singularities. As these regions have smaller Fermi velocities, this leads to a reduction of the mean-squared velocities determining the field scales $H_{i0}$ in Eq. (2) and, consequently, to an increase of the upper critical fields at all temperatures.

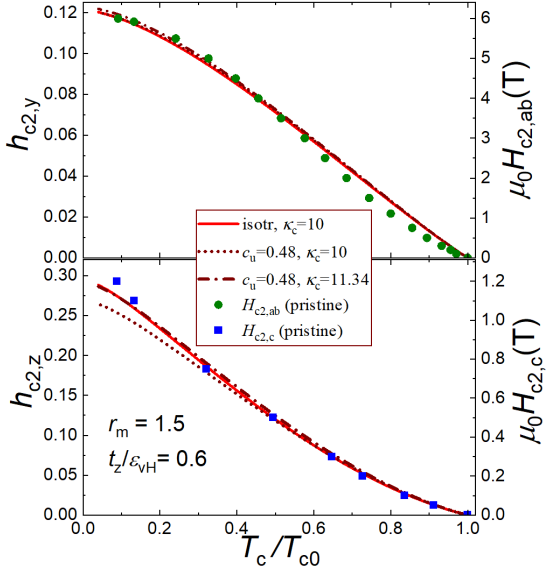

FIG. S9. The reduced upper critical fields for representative parameters illustrating the role of gap anisotropy controlled by the parameters $c_u$ and $c_v$ in Eq. (B23). We assumed the relation $r_m c_v = c_u$.

For the $c$-axis orientation, the relative increase of $H_{c2}$ near $T_c$ is stronger than at low temperatures, thereby reducing the upward curvature. Figure S9 illustrates the influence of finite gap anisotropy on the $H_{c2}$ shape for two field orientations using representative parameters. The choice of parameters corresponds to large gap anisotropy, $\Delta_{min}/\Delta_{max} = 0.2$. We see that for $c$-axis direction the upward curvature decreases, while for the in-plane direction it remains practically unchanged. One can make the upward curvature close to the isotropic case by increasing the parameter $\kappa_c$ from 10 to 11.34. This correction increases the upward curvature for the in-plane orientation with respect to the isotropic case.

**References**

1. B. R. Ortiz, S. M. L. Teicher, Y. Hu, J. L. Zuo, P. M. Sarte, *et al*., CsV$_3$Sb$_5$: A Z$_2$ Topological Kagome metal with a superconducting ground state, *Phys. Rev. Lett*. **125**, 247002 (2020).
2. Y. Fu, N. Zhao, Z. Chen, Q. Yin, Z. Tu, C. Gong, C. Xi, X. Zhu, Y. Sun, K. Liu, and H. Lei, Quantum transport evidence of topological band structures of Kagome superconductor CsV$_3$Sb$_5$, *Phys. Rev. Lett.* **127**, 207002 (2021).
3. F. H. Yu, T. Wu, Z. Y. Wang, B. Lei, W. Z. Zhuo, J. J. Ying, and X. H. Chen, Concurrence of anomalous Hall effect and charge density wave in a superconducting topological kagome metal, *Phys. Rev. B* **104**, L041103 (2021).
4. J. F. Ziegler, J. P. Biersack, and M. D. Ziegler. SRIM, The Stopping and range of Ions in Matter, Ion Implantation Press (2008). Available at: www.lulu.com/content/1524197
5. A. A. Abrikosov and L. P. Gorkov, On the theory of superconducting alloys, I. The electrodynamics of alloys at absolute zero, *Zh. Eksp. Teor. Fiz*. **35**, 1558 (1958) [*Sov. Phys. JETP* **8**, 1090 (1959)].
6. L. A. Openov, Critical temperature of an anisotropic superconductor containing both nonmagnetic and magnetic impurities, *Phys. Rev. B* **58**, 9468 (1998).
7. E. I. Timmons, S. Teknowijoyo, M. Konczykowski, O. Cavani, M. A. Tanatar, S. Ghimire, K. Cho, Y. Lee, L. Ke, N. H. Jo *et al*, Electron irradiation effects on superconductivity in PdTe$_2$: An application of a generalized Anderson theorem, *Phys. Rev. Res*. **2**, 023140 (2020).
8. J. Liu, Q. Li, Y. Li, X. Fan, J. Li, P. Zhu, H. Deng, J.-X.Yin, H. Yang, J. Li, and H. -H. Wen, Enhancement of superconductivity and phase diagram of Ta-doped Kagome superconductor CsV$_3$Sb$_5$, *Sci. Rep*. **14**, 9580. (2024).
9. S. L. Ni, S. Ma, Y. H. Zhang, J. Yuan, H. T. Yang, Z. Y. W. Lu, N. N. Wang, J. P. Sun, Z. Zhao, D. Li et al, Anisotropic superconducting properties of kagome metal CsV$_3$Sb$_5$, *Chin. Phys. Lett*. **38**, 057403 (2021).
10. W. Duan, Z. Nie, S. Luo, F. Yu, B. R. Ortiz, L. Yin, H. Su, F. Du, A. Wang, Y. Chen, X. Lu, J. Ying, S. D. Wilson, X. Chen, Y. Song, and H. Yuan, Nodeless superconductivity in the kagome metal CsV$_3$Sb$_5$, *Sci. China-Phys. Mech. Astron*. **64**, 107462 (2021).
11. P. W. Anderson, Theory of dirty superconductors, *J. Phys. Chem. Solids*, **11**, 26 (1959).
12. E. I. Timmons, S. Teknowijoyo, M. Konczykowski, O. Cavani, M. A. Tanatar, S. Ghimire, K. Cho, Y. Lee, L. Ke, N. H. Jo *et al*, Electron irradiation effects on superconductivity in PdTe$_2$: An application of a generalized Anderson theorem, *Phys. Rev. Res*. **2**, 023140 (2020).
13. A. A. Abrikosov and L. P. Gorkov, On the theory of superconducting alloys, I. The electrodynamics of alloys at absolute zero, *Zh. Eksp. Teor. Fiz*. **35**, 1558 (1958) [*Sov. Phys. JETP* **8**, 1090 (1959)].
14. M. Roppongi, K. Ishihara, Y. Tanaka, K. Ogawa, K. Okada, S. Liu, K. Mukasa, Y. Mizukami, Y. Uwatoko, R. Grasset, M. Konczykowski, B. R. Ortiz, S. D. Wilson, K. Hashimoto, and T. Shibauchi, Bulk evidence of anisotropic s-wave pairing with no sign change in the kagome superconductor CsV$_3$Sb$_5$, *Nat. Commun*. **14**, 667 (2023).

15. P. Hohenberg, Anisotropic superconductors with nonmagnetic impurities, *Zh. Eksp. Teor. Fiz.* **45**, 1208 (1963) (Sov. Phys.-JETP 18, 834 (1964)].

16. L. A. Openov, Critical temperature of an anisotropic superconductor containing both nonmagnetic and magnetic impurities, *Phys. Rev. B* **58**, 9468 (1998).

17. K. Fukushima, K. Obata, S. Yamane, Y. Hu, Y. Li, Y. Yao, Z. Wang, Y. Maeno, S. Yonezawa, Violation of emergent rotational symmetry in the hexagonal Kagome superconductor $CsV_3Sb_5$, *Nat. Commun*. **15**, 2888 (2024).

18. L. Nie, K. Sun, W. Ma, D. Song, L. Zheng, Z. Liang, P. Wu, F. Yu, J. Li, M. Shan, D. Zhao, S. Li, B. Kang, Z. Wu, Y. Zhou, K. Liu, Z. Xiang, J. Ying, Z. Wang, T. Wu, X. Chen, Charge-density-wave driven nematicity in Kagome superconductor, *Nature* **604**, 59 (2022).

19. F. C. Menegotto, R. S. Severino, P. D. Mininni, E. Fradkin, V. Bekeris, G. Pasquini, G. S. Lozano, Vortex flow anisotropy in nematic superconductors, arXiv:2501.21794.

20. J. Plumb, A. C. Salinas, K. Mallayya, E. Kisiel, F. B. Carneiro, R. Gomez, G. Pokharel, E.-A. Kim, S. Sarker, Z. Islam, S. Daly, S. D. Wilson, Phase-separated charge order and twinning across length scales in $CsV_3Sb_5$, *Phys. Rev. Materials* **8**, 093601 (2024).

21. D. Subries, A. Korshunov, A. H. Said, L. Sanchez, B. R. Ortiz, S. D. Wilson, A. Bosak, S. Blanco-Canosa, Order-disorder charge density wave instability in the Kagome metal $(Cs,Rb)V_3Sb_5$, *Nat. Commun*. **14**, 1015 (2023).

22. L. Kautsch, Y. M. Oey, H. Li, Z. Ren, B. R. Ortiz, G. Pokharel, R. Seshadri, J. Ruff, T. Kongruengkit, J. W. Harter, Z. Wang, I. Zelkovic, S. D. Wilson, Incommensurate charge-stripe correlations in the kagome superconductor $CsV_3Sb_{5-x}Sn_x$, *npj Quantum Materials* **8**, 37 (2023).

23. H. Zhao, H. Li, B. R. Ortiz, S. M. Teicher, T. Park, M. Ye, Z. Wang, L. Balents, S. D. Wilson, I. Zelkovic, Cascade of correlated electron states in the Kagome superconductor $CsV_3Sb_5$, *Nature* **599**, 216 (2021).

24. Y. Xu, Z. Ni, Y. Liu, B. R. Ortiz, Q. Deng, S. D. Wilson, B. Yan, L. Balents, L. Wu, Three-state nematicity and magneto-optical Kerr effect in the charge density waves in Kagome superconductors, *Nat. Phys*. **18**, 1470 (2022).

25. N. R. Werthamer, E. Helfand, and P. C. Hohenberg, Temperature and purity dependence of the superconducting critical field, $H_{c2}$. III. Electron spin and spin-orbit Effects, *Phys. Rev*. **147**, 295 (1966).

26. V. G. Kogan and R. Prozorov, Orbital upper critical field and its anisotropy of clean one-and two-band superconductors, *Rep. Prog. Phys*. **75**, 114502 (2012).

27. R. G. Dias and J. M. Wheatley, Superconducting upper critical field near a 2D van Hove singularity, *Solid State Commun.*, **98**, 859 (1996).

28. R. O. Zaitsev, On the effect of van Hove singularities on the critical field of type-II superconductors, *JETP Letters*, **65**, 74 (1997). (Pis'ma Zh. Eksp. Teor. Fiz. **65**, 71 (1997)).

29. R. G. Dias, Effects of van Hove singularities on the upper critical field, *J. Phys.: Condens. Matter*. **12**, 9053 (2000).

30. A. E. Koshelev, R. Chapai, D. Y. Chung, J. F. Mitchell, and U. Welp, Origin of anomalous magnetotransport in kagome superconductors $AV_3Sb_5$ (A = K, Rb, Cs), *Phys. Rev. B* **110**, 024512 (2024).